\documentclass[12pt]{iopart}

\usepackage{iopams}
\usepackage{graphicx}
\usepackage{mathrsfs}
\usepackage{color}
\usepackage[table]{xcolor}
\usepackage{hyperref}
\hypersetup{colorlinks, citecolor=black, filecolor=black, linkcolor=black, urlcolor=black}

\expandafter\let\csname equation*\endcsname\relax
\expandafter\let\csname endequation*\endcsname\relax
\usepackage{amsmath}
\allowdisplaybreaks[1]

\definecolor{grayish}{RGB}{230,230,230}

\newcommand{\refEq}[1] {(\ref{#1})}

\newcommand{\Sin}[1]{\ensuremath{\sin \left( #1 \right)}}
\newcommand{\Cos}[1]{\ensuremath{\cos \left( #1 \right)}}

\newcommand{\ArcTan}[1]{\ensuremath{\text{arctan} \left( #1 \right)}}

\newcommand{\Exp}[1]{\ensuremath{\exp \left( #1 \right)}}

\newcommand{\Nabla}{\ensuremath{\vec{\nabla}}}

\newcommand{\Order}[1]{\ensuremath{O \left( #1 \right)}}
\newcommand{\romanNum}[1]{\uppercase\expandafter{\romannumeral#1}}

\begin{document}

\title[Momentum transport due to the beating between flux surface shaping effects]{Turbulent momentum transport due to the beating between different tokamak flux surface shaping effects}

\author{Justin Ball and Felix I. Parra}

\address{Rudolf Peierls Centre for Theoretical Physics, University of Oxford, Oxford OX1 3NP, United Kingdom}
\address{Culham Centre for Fusion Energy, Culham Science Centre, Abingdon OX14 3DB, United Kingdom}
\ead{Justin.Ball@physics.ox.ac.uk}

\begin{abstract}

Introducing up-down asymmetry into the tokamak magnetic equilibria appears to be a feasible method to drive fast intrinsic toroidal rotation in future large devices. In this paper we investigate how the intrinsic momentum transport generated by up-down asymmetric shaping scales with the mode number of the shaping effects. Making use the gyrokinetic tilting symmetry (Ball {\it et al} (2016) {\it Plasma Phys. Control. Fusion} {\bf 58} 045023), we study the effect of envelopes created by the beating of different high-order shaping effects. This reveals that the presence of an envelope can change the scaling of the momentum flux from exponentially small in the limit of large shaping mode number to just polynomially small. This enhancement of the momentum transport requires the envelope to be both up-down asymmetric and have a spatial scale on the order of the minor radius.

\end{abstract}



\section{Introduction}
\label{sec:introduction}

Driving toroidal plasma rotation in tokamaks is important for many reasons. The absolute magnitude of the rotation can stabilize resistive wall modes \cite{LiuITERrwmStabilization2004, OkabayashiActiveRWMfeedback2001, GarofaloActiveRWMfeedback2001}, a class of MHD instabilities that can cause disruptions. Gradients in the rotation can directly reduce turbulence, thereby increasing the energy confinement time \cite{RitzRotShearTurbSuppression1990, BurrellShearTurbStabilization1997, BarnesFlowShear2011, HighcockRotationBifurcation2010}. Furthermore, edge rotation is thought to enable the transition to H-mode \cite{WagnerHmodeReview2007}, an improved confinement regime that is desirable for tokamak fusion reactors \cite{TakizukaHmodeThresholdDatabase2004}.

Tokamak plasmas will start to rotate if pushed using externally-injected momentum. This is commonly done with beams of neutral particles and is usually the dominant drive of rotation in current experiments. However, future devices (e.g. the ITER experiment \cite{AymarITERSummary2001} or a demonstration power plant) are expected to have much larger plasmas than current experiments. Larger plasmas have significantly more inertia and require higher velocity neutral beams in order to penetrate to the plasma center. Because energy is quadratic with velocity and momentum is linear, the ratio of the momentum to energy carried by a neutral beam varies inversely with the beam velocity \cite{ParraMomentumTransitions2011}. Hence, the neutral beams used to heat larger devices are expected to be less effective at driving rotation \cite{LiuITERrwmStabilization2004}.

\begin{figure}
    \hspace{0.04\textwidth} (a) \hspace{0.255\textwidth} (b) \hspace{0.255\textwidth} (c) \hspace{0.1\textwidth}
	\begin{center}
		\includegraphics[height=0.33\textwidth]{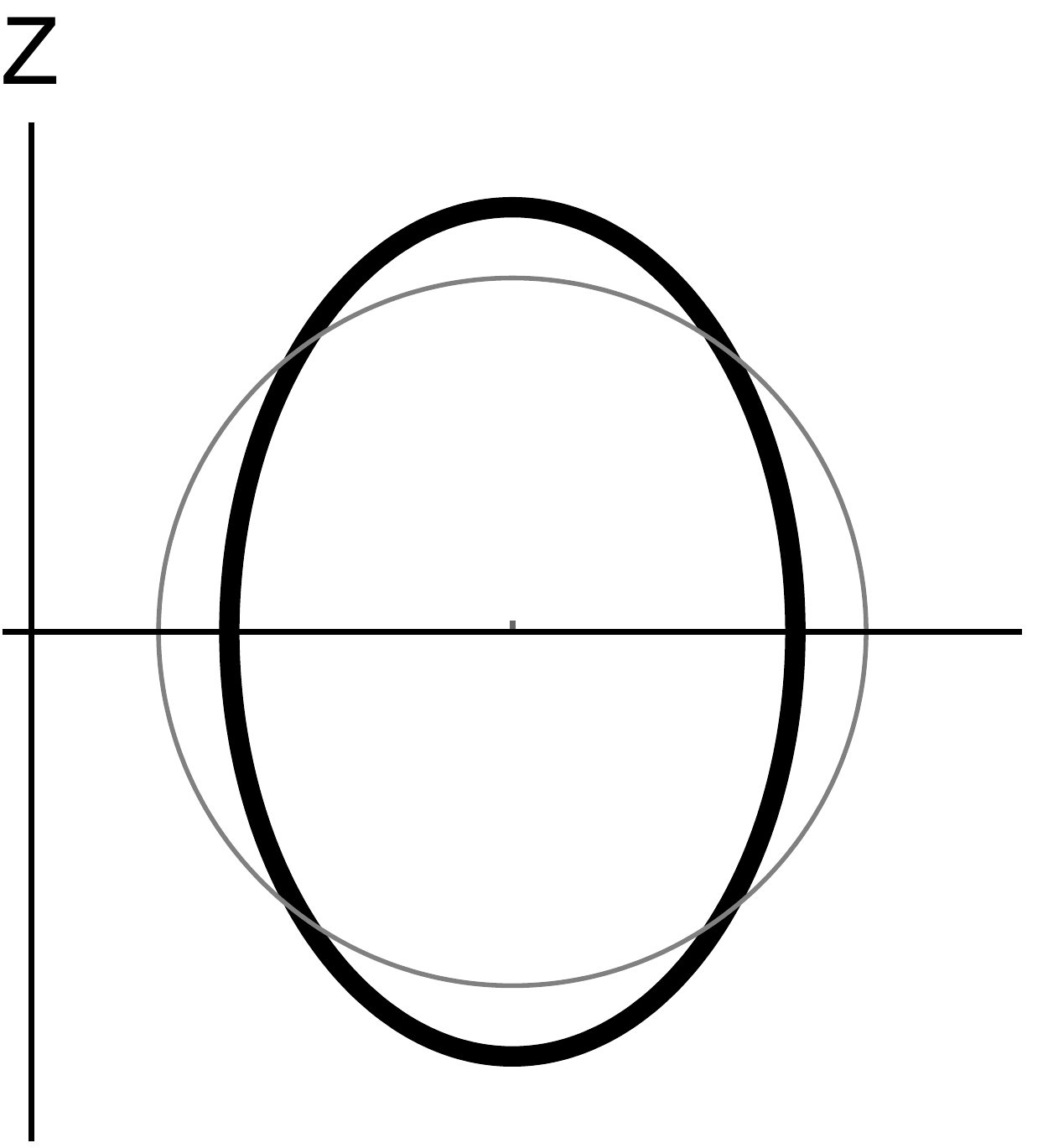}
		\includegraphics[height=0.33\textwidth]{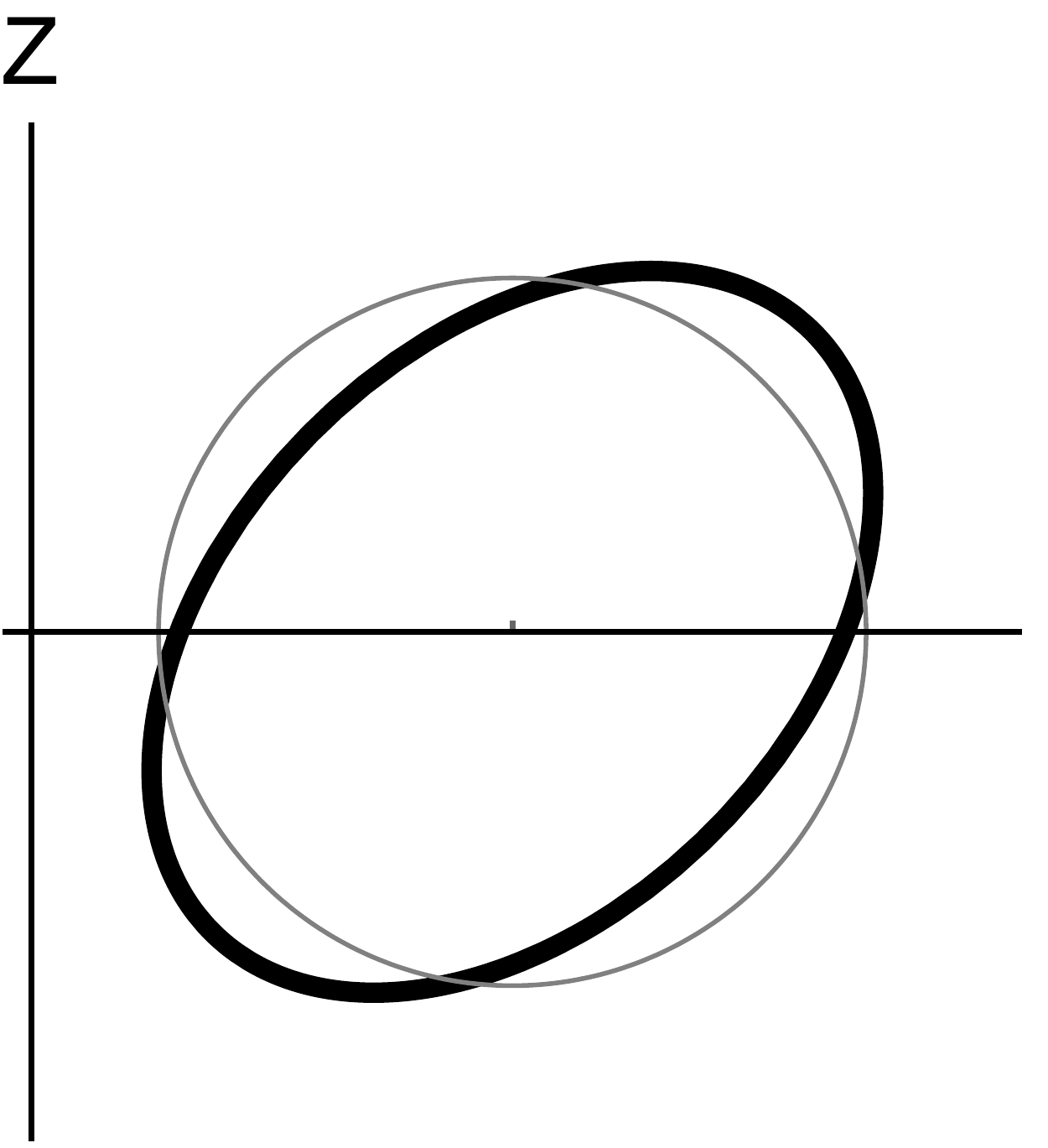}
		\includegraphics[height=0.33\textwidth]{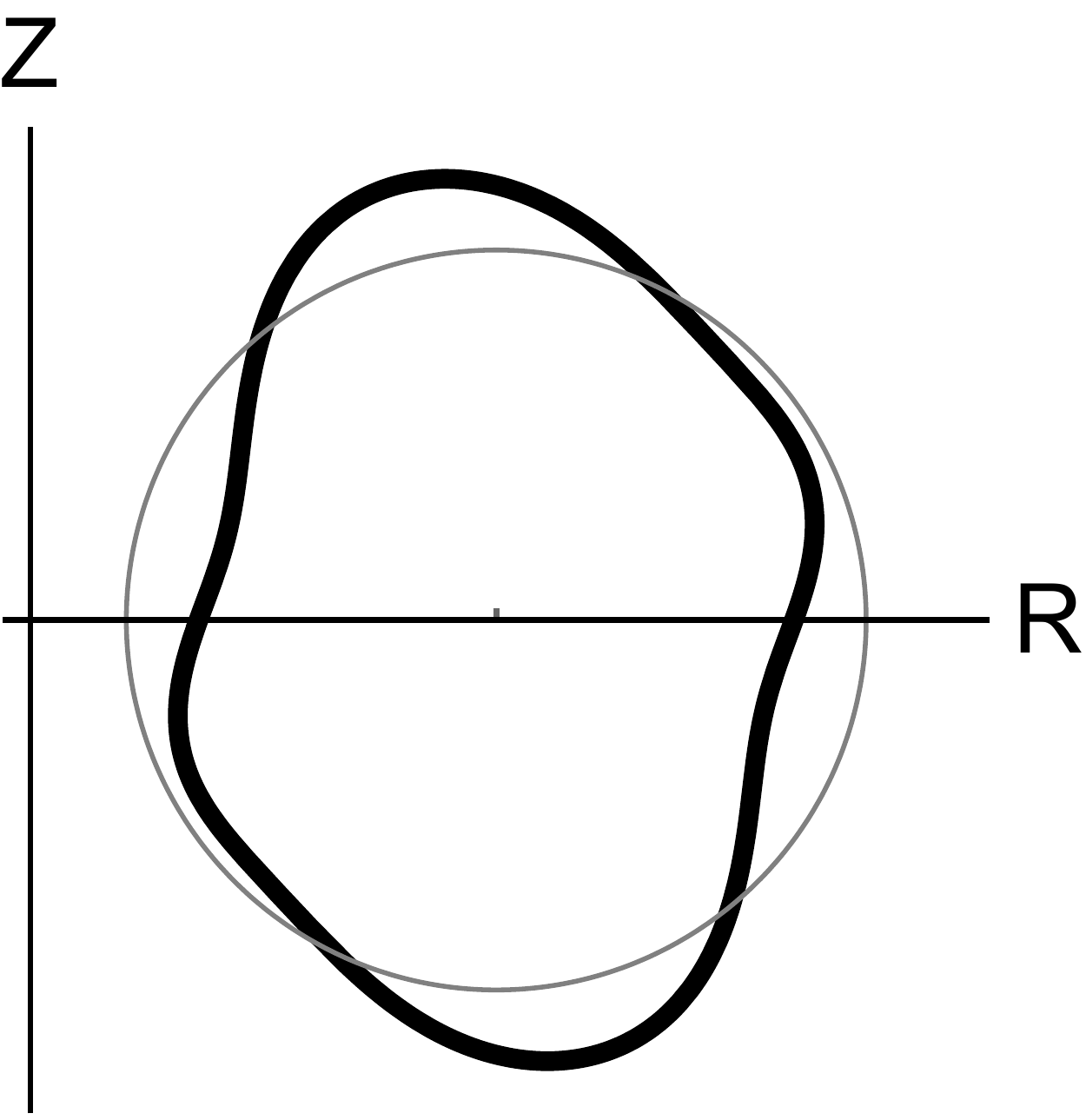}
	\end{center}
	\caption{Cartoon flux surfaces (black) with no beating envelopes that are (a) up-down symmetric, (b) mirror symmetric, but up-down asymmetric, and (c) non-mirror symmetric, where the horizontal axis is the midplane and circular flux surfaces (gray) are shown for comparison.}
	\label{fig:updownSymShaping}
\end{figure}

An alternative is ``intrinsic'' rotation, or spontaneous rotation observed in the absence of external injection \cite{RiceExpIntrinsicRotMeas2007}. This rotation arises from plasma turbulence moving momentum between tokamak flux surfaces and is especially attractive because it does not require any external power. However, the mechanisms driving intrinsic rotation in the core of current experiments are expected to diminish in larger devices. This is because of the up-down symmetry of the lowest-order local $\delta f$ gyrokinetic equation \cite{PeetersMomTransSym2005, ParraUpDownSym2011, SugamaUpDownSym2011}, which is thought to accurately model turbulence in the core. This symmetry implies that tokamak turbulence (when averaged over the turbulent timescale) generates momentum flux that has odd parity about the midplane (the midplane is shown in figure \ref{fig:updownSymShaping}). Hence, the net momentum flux across a flux surface is zero to the accuracy of lowest-order gyrokinetics, meaning that intrinsic rotation must small in $\rho_{\ast} \equiv \rho_{i} / a \ll 1$ (i.e. the ratio of the ion gyroradius to the tokamak minor radius). Since large machines have smaller values of $\rho_{\ast}$, the level of intrinsic rotation is expected to decrease. 

However, there is one mechanism capable of breaking the symmetry of the turbulence to generate lowest order rotation in a stationary plasma: up-down asymmetric plasma shaping. When the tokamak flux surfaces do not have mirror symmetry about the midplane (e.g. figure \ref{fig:updownSymShaping}(b,c)), the momentum transport above the midplane is no longer guaranteed to cancel the momentum transport below it. Hence, large toroidal flows can spontaneously develop. In fact, reference \cite{CamenenPRLExp2010} presents results from the TCV tokamak that have provided the first experimental evidence of intrinsic rotation generated by up-down asymmetry. Subsequently, reference \cite{BallMomUpDownAsym2014} presents nonlinear gyrokinetic simulations that are consistent with the TCV results and suggest that up-down asymmetry is a feasible method to generate the current, experimentally-measured rotation levels in power plant-sized devices.

Building on this work, references \cite{BallMirrorSymArg2016} and \cite{BallMomFluxScaling2016} use gyrokinetics to argue that certain types of flux surface shapes may increase the intrinsic rotation driven by up-down asymmetry. To make the problem analytically tractable, both explore breaking up-down symmetry using ``fast'' shaping effects, where ``fast'' refers to shaping with a small spatial scale (i.e. shaping with a characteristic poloidal mode number $m_{c} \gg 1$). In this limit, reference \cite{BallMirrorSymArg2016} demonstrates another gyrokinetic symmetry, namely poloidal tilting symmetry. This symmetry implies that a poloidal translation of all fast poloidal variation (i.e. that of order $m_{c}$) by a single tilt angle has an exponentially small effect in $m_{c} \gg 1$ on the turbulent transport.

Using this poloidal tilting symmetry and the up-down symmetry mentioned above, reference \cite{BallMomFluxScaling2016} distinguishes flux surfaces with mirror symmetry (e.g. figure \ref{fig:updownSymShaping}(b)) from non-mirror symmetric flux surfaces (e.g. figure \ref{fig:updownSymShaping}(c)). From the up-down symmetry argument, we know that up-down symmetric flux surfaces (e.g. figure \ref{fig:updownSymShaping}(a)) generate no momentum flux in the gyrokinetic model. Since, by definition, mirror symmetric tokamaks must have mirror symmetry about some line in the poloidal plane, we can rotate all of the shaping effects by a single tilt angle until the line of mirror symmetry is coincident with the midplane. Hence, by the tilting symmetry argument, we know that the momentum flux in mirror symmetric tokamaks is exponentially small in $m_{c} \gg 1$. This argument only relies on an expansion in $m_{c} \gg 1$ to distinguish the fast poloidal variation from slow variation. It does \textit{not} presume that the flux surface shaping is weak.

\emph{In this work, we investigate a property of the momentum transport that went unexamined in the analysis of \cite{BallMirrorSymArg2016} and \cite{BallMomFluxScaling2016}: the impact of a slowly-varying envelope created by the beating of two or more fast flux surface shaping effects.} This work will use much of the analysis from \cite{BallMirrorSymArg2016} and \cite{BallMomFluxScaling2016}, but will explicitly study momentum transport generated by flux surfaces with slowly-varying envelopes. We will also provide an overview of the expected scaling of momentum flux with shaping mode number for all flux surface shapes.

In section \ref{sec:analytics} we present the analytic arguments that are needed to understand the effect of an envelope on the momentum transport. In section \ref{subsec:model} we detail the analytic models we use to both specify the magnetic equilibrium and calculate the turbulent momentum flux. Next, in section \ref{subsec:tiltingSym}, we revisit the tilting symmetry of \cite{BallMirrorSymArg2016} in order to determine when flux surfaces with envelopes must have exponentially small momentum transport. Subsequently, in section \ref{subsec:example} we clarify what we have learned by considering a few example flux surfaces. Then, we revisit the calculations of \cite{BallMomFluxScaling2016} in order to show that we expect the momentum transport to be polynomially small when the tilting symmetry does not constrain it to be exponentially small. These analytic arguments are then compared against numerical simulations in section \ref{sec:numerics}. Lastly, section \ref{sec:conclusions} provides some concluding remarks.

\section{Analytic gyrokinetic analysis}
\label{sec:analytics}

\subsection{Analytic model}
\label{subsec:model}

In order to model the transport of momentum in the core of tokamaks we will use gyrokinetics \cite{LeeGenFreqGyro1983, LeeParticleSimGyro1983, DubinHamiltonianGyro1983, HahmGyrokinetics1988, SugamaGyroTransport1996, SugamaHighFlowGyro1998, BrizardGyroFoundations2007, ParraGyrokineticLimitations2008, ParraLagrangianGyro2011, AbelGyrokineticsDeriv2012} because experimental measurements \cite{McKeeTurbulenceScale2001} indicate that it accurately treats turbulence. Gyrokinetics is a fully kinetic description based on an expansion of the Fokker-Planck and Maxwell's equations in $\rho_{\ast} \ll 1$. It specifically investigates behavior much slower than the ion gyrofrequency, but still allows the size of the turbulence perpendicular to the magnetic field to be comparable to the gyroradius. In this regime, we can average over the fast gyromotion of particles. This removes one dimension of velocity space as well as the gyrofrequency timescale, which makes the model computationally tractable.

In keeping with \cite{BallMomFluxScaling2016}, we will use $\delta f$ electrostatic gyrokinetics to study turbulence in the local vicinity of a single field line on a single flux surface of interest. In doing so we expand the distribution function in $\rho_{\ast} \ll 1$, assuming that the lowest order contribution is Maxwellian, and calculate the perturbation. Note that we will neglect pre-existing background rotation because we are interested in driving rotation in a stationary plasma. With these assumptions we can solve the Fourier-analyzed gyrokinetic equation (e.g. equation (2) of \cite{BallMomFluxScaling2016}) with the Fourier-analyzed quasineutrality equation (e.g. equation (11) of \cite{BallMomFluxScaling2016}) to find the the perturbed distribution function and electrostatic potential. This allows us to calculate the radial flux of toroidal angular momentum from equation (13) of \cite{BallMomFluxScaling2016}.

For this work the eight geometric coefficients that appear in the gyrokinetic model are particularly important. They are $B$, $\hat{b} \cdot \Nabla \theta$, $\vec{v}_{d s} \cdot \Nabla \psi$, $\vec{v}_{d s} \cdot \Nabla \alpha$, $a_{s ||}$, $\left| \Nabla \psi \right|^{2}$, $\Nabla \psi \cdot \Nabla \alpha$, and $\left| \Nabla \alpha \right|^{2}$. Here $\vec{B}$ is the magnetic field, $\hat{b} \equiv \vec{B} / B$ is the magnetic field unit vector, $\theta$ is the usual cylindrical poloidal angle measured from the midplane, $\vec{v}_{d s}$ is the magnetic drift velocity, $\psi$ is the poloidal magnetic flux,
\begin{align}
   \alpha \equiv \zeta - \int_{\theta_{\alpha} \left( \psi \right)}^{\theta} d \theta' \frac{I \left( \psi \right)}{R^{2} \vec{B} \cdot \Nabla \theta'} \label{eq:alphaDef}
\end{align}
is a coordinate that selects a particular field line from a given flux surface, $I \left( \psi \right) \equiv R B_{\zeta}$ is the toroidal field flux function, $R$ is the major radial coordinate, $\zeta$ is the toroidal angle, $\theta_{\alpha} \left( \psi \right)$ is a free function that determines the field line selected by $\alpha = 0$, and $a_{s ||}$ is the parallel acceleration. These coefficients are the only way the magnetic geometry enters into gyrokinetics, so they are critical in understanding intrinsic rotation driven by up-down asymmetry.

The eight coefficients are each calculated from the tokamak equilibrium, which we will specify using a generalization of the Miller local equilibrium model \cite{MillerGeometry1998} (see section 2.1 of \cite{BallMomFluxScaling2016} for details). The Miller model requires the shape of the flux surface of interest, how the shape changes with minor radius, and a number of scalar quantities. To keep the specification as general as possible, we will prescribe the flux surface shape and its radial gradient using a Fourier decomposition as
\begin{align}
   r_{0} \left( \theta \right) &= r_{\psi 0} \left( 1 - \sum_{m = 1}^{\infty} \frac{\Delta_{m} - 1}{\Delta_{m} + 1} \Cos{m \left( \theta + \theta_{t m} \right)} \right) \label{eq:fluxSurfaceSpec} \\
   \left. \frac{\partial r}{\partial r_{\psi}} \right|_{r_{\psi 0}} &= 1 - \sum_{m = 1}^{\infty} \left( \frac{\Delta_{m} - 1}{\Delta_{m} + 1} + \frac{2 r_{\psi 0}}{\left( \Delta_{m} + 1 \right)^{2}} \frac{d \Delta_{m}}{d r_{\psi}} \right) \Cos{m \left( \theta + \theta_{t m} \right)} \label{eq:fluxSurfaceChangeSpec}
\end{align}
respectively. Here $r$ is the usual cylindrical minor radius, $r_{0}$ is the minor radius of the flux surface of interest (which varies with poloidal angle), $r_{\psi}$ is the minor radial flux surface label, $r_{\psi 0}$ is the value of $r_{\psi}$ on the flux surface of interest, $m$ is the poloidal shaping mode number, $\Delta_{m}$ indicates the magnitude of each shaping effect, and $\theta_{t m}$ is the tilt angle of each shaping effect (defined such that an increasing $\theta_{t m}$ rotates the shaping in the direction of decreasing $\theta$). The quantity $\Delta_{m}$ is the ratio of the maximum value of the minor radius to the minimum value of the minor radius on the flux surface, if the mode $m$ was the only shaping effect present. Hence, we find that the $m=2$ mode approximately corresponds to ellipticity and $\Delta_{2}$ is identical to $\kappa$, the usual flux surface elongation. Similarly, the $m=1$ mode controls the Cartesian translation of the flux surface shape (i.e. the Shafranov shift), though it also modifies the flux surface shape to ensure that the ratio of the maximum to minimum minor radius corresponds to $\Delta_{1}$. We can use equations \refEq{eq:fluxSurfaceSpec} and \refEq{eq:fluxSurfaceChangeSpec} together with
\begin{align}
   r \left( r_{\psi}, \theta \right) =& r_{0} \left( \theta \right) + \left. \frac{\partial r}{\partial r_{\psi}} \right|_{\psi_{0}} \left( r_{\psi} - r_{\psi 0} \right) \label{eq:gradShafranovLocalEq} \\
   R \left( r_{\psi}, \theta \right) =& R_{0} + r \left( r_{\psi}, \theta \right) \Cos{\theta} \label{eq:geoMajorRadial} \\
   Z \left( r_{\psi}, \theta \right) =& r \left( r_{\psi}, \theta \right) \Sin{\theta} , \label{eq:geoAxial}
\end{align}
to fully specify the local flux surface geometry of the tokamak equilibrium. Here $R_{0}$ is the major radial location of the flux surface of interest and $Z$ is the axial coordinate.

This completes the theoretical model. We can specify the magnetic equilibrium, which allows us to calculate the eight geometric coefficients appearing in gyrokinetics. Then, from gyrokinetics, we calculate the distribution function and electrostatic potential on the flux surface of interest. This determines the momentum flux, which we want to maximize by specifying the optimal up-down asymmetric flux surface shape.

\subsection{Tilting symmetry}
\label{subsec:tiltingSym}

\begin{figure}
    \hspace{0.04\textwidth} (a) \hspace{0.255\textwidth} (b) \hspace{0.255\textwidth} (c) \hspace{0.1\textwidth}
	\begin{center}
		\includegraphics[height=0.33\textwidth]{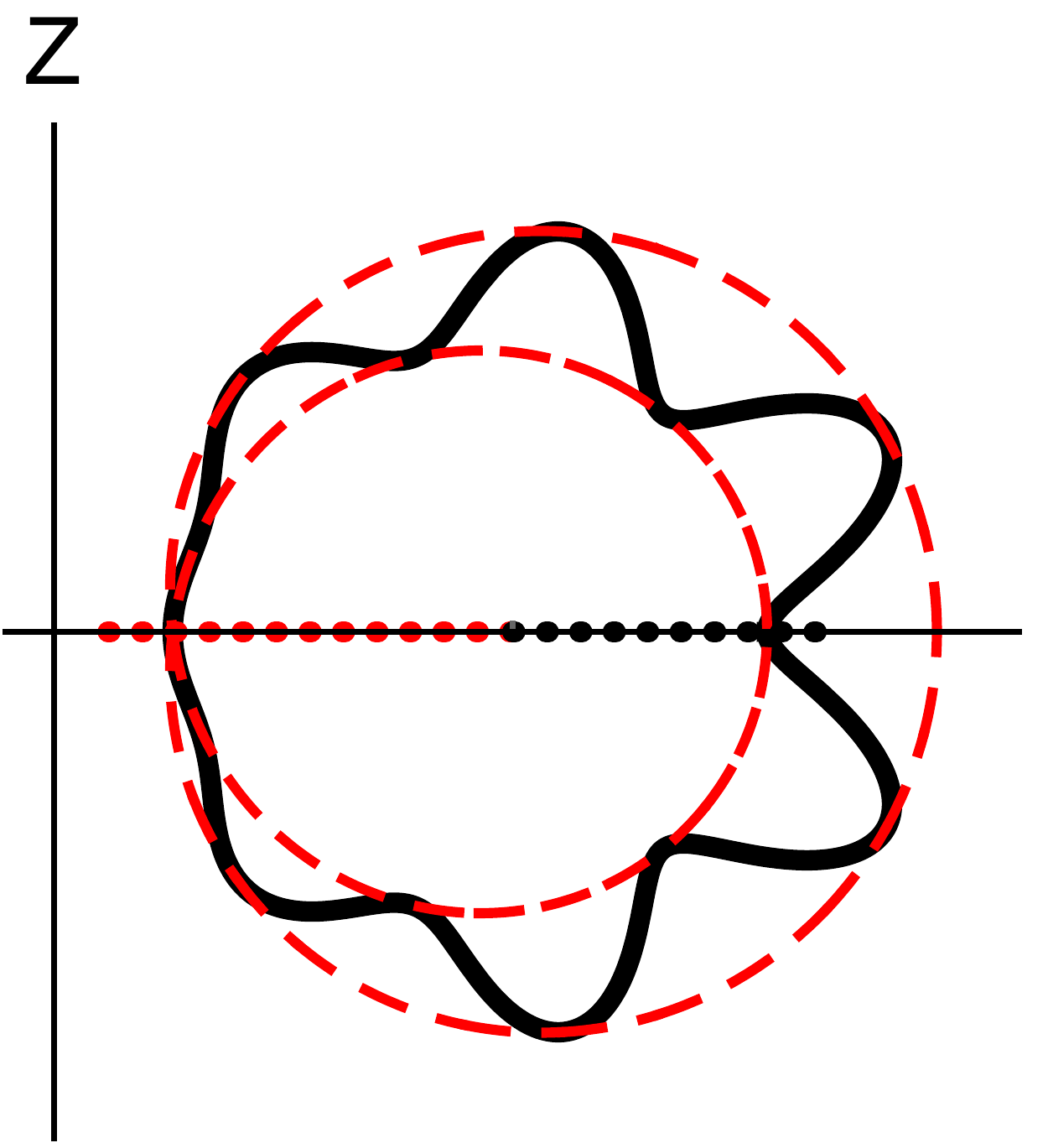}
		\includegraphics[height=0.33\textwidth]{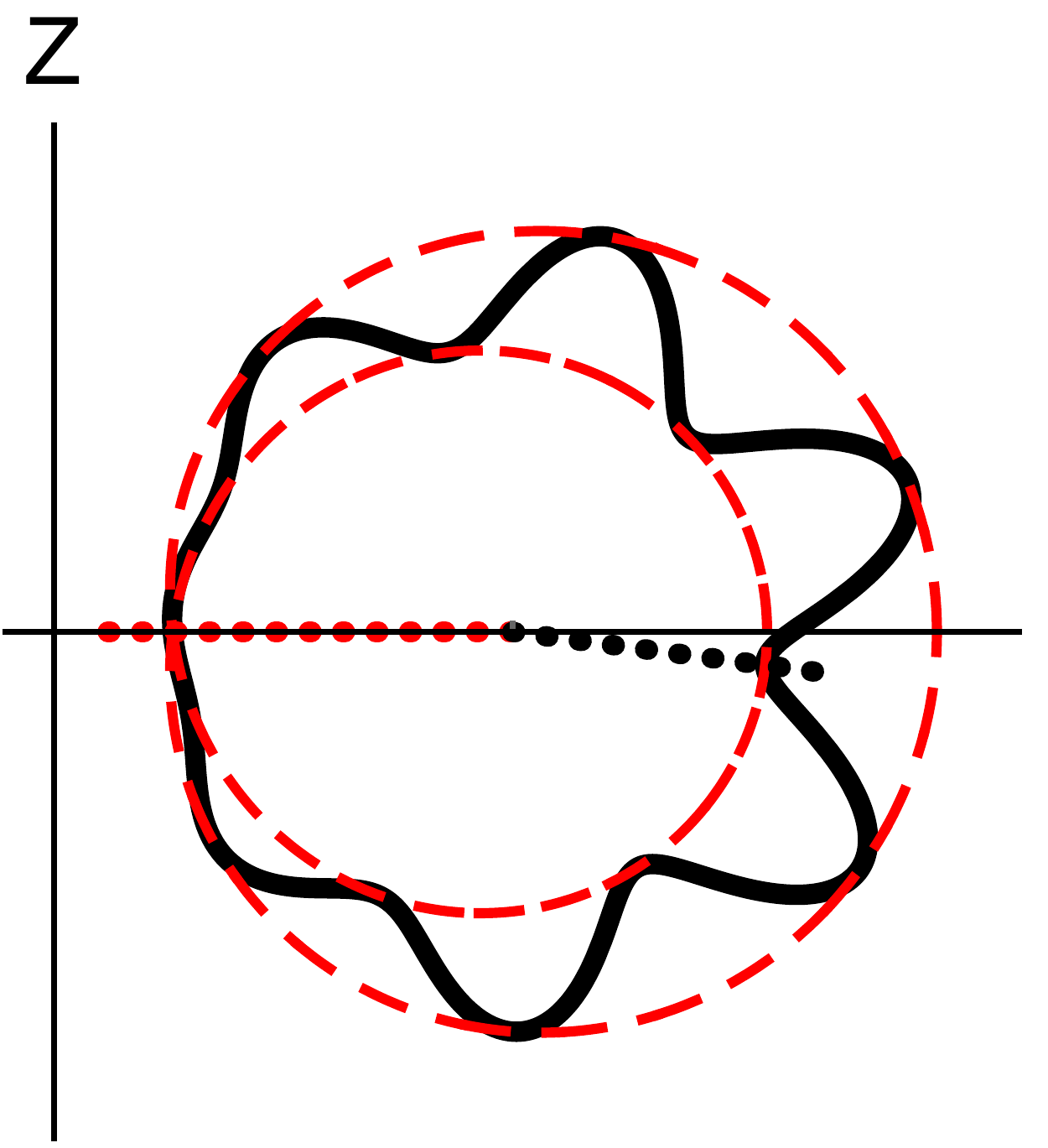}
		\includegraphics[height=0.33\textwidth]{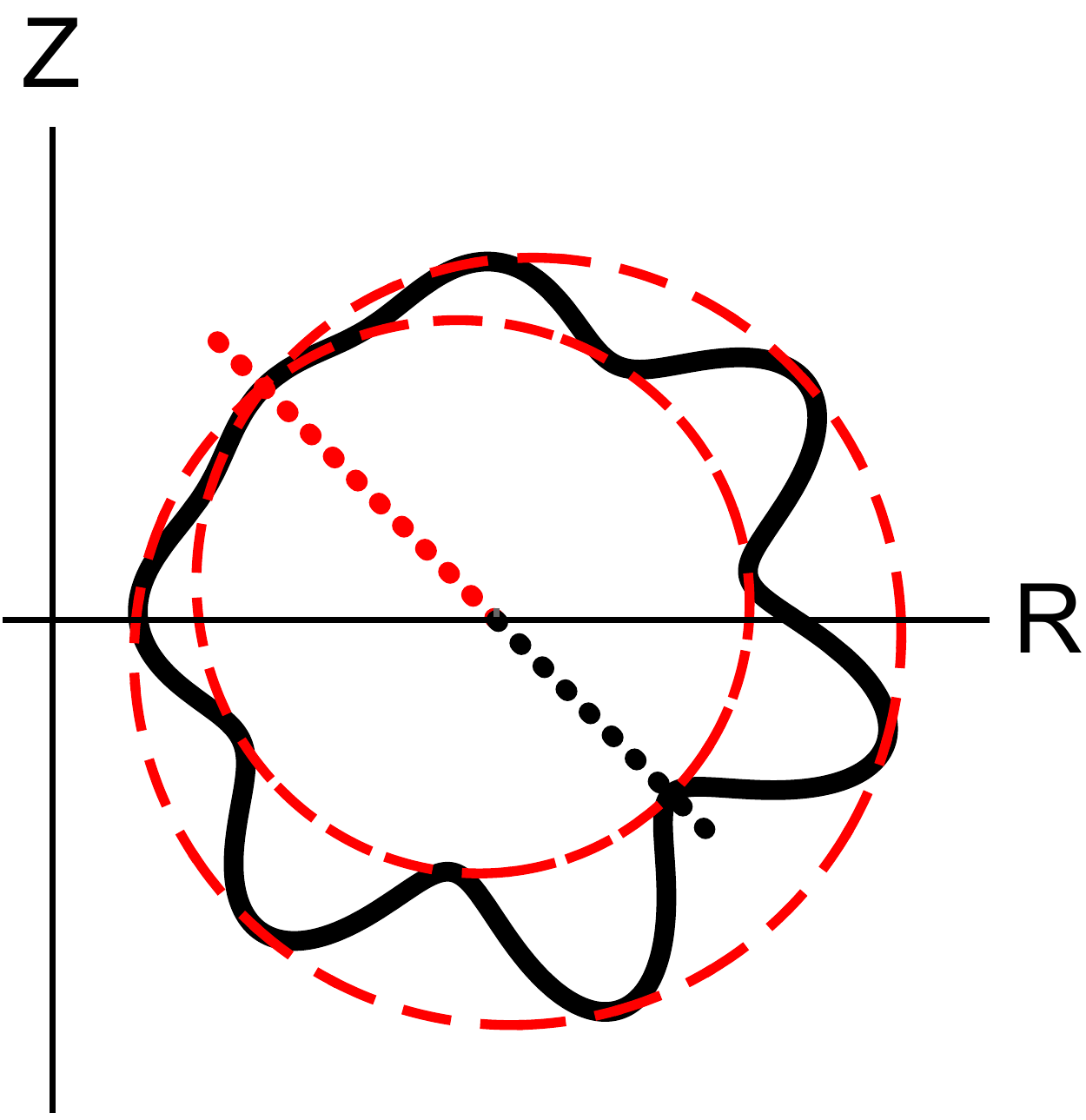}
	\end{center}
	\caption{Cartoon flux surfaces (black, solid) with an envelope (red, dashed) and (a) no tilt, (b) a $\pi/4$ tilt of the fast poloidal variation, and (c) a $\pi/4$ tilt of the fast shaping effects, where the tilt of the fast shaping (black, dotted) and the envelope (red, dotted) are indicated.}
	\label{fig:envelopeShaping}
\end{figure}

When considering a slowly-varying envelope it becomes important to distinguish ``fast shaping effects'' from ``fast poloidal variation.'' ``Fast shaping effects'' refer to the terms that have $m \gg 1$ in the Fourier decomposition of the flux surface shape (i.e. equation \refEq{eq:fluxSurfaceSpec}). ``Fast poloidal variation'' on the other hand refers to variation in the flux surface shape that has a spatial scale much smaller than the minor radius. This distinction is important because tilting all shaping effects with large poloidal mode number also tilts any envelope, while tilting only the fast poloidal variation keeps any slowly-varying envelopes fixed. Figure \ref{fig:envelopeShaping}(a) shows a cartoon of an up-down symmetric flux surface, which can become figure \ref{fig:envelopeShaping}(b) after a tilt of only the fast poloidal variation or figure \ref{fig:envelopeShaping}(c) after a tilt of the fast shaping effects.

In order to make use of the tilting symmetry of \cite{BallMirrorSymArg2016} we must first introduce a fast poloidal coordinate
\begin{align}
z \equiv m_{c} \theta \label{eq:zDef}
\end{align}
and separate the two poloidal scales. Here $m_{c}$ is a characteristic mode number that indicates the boundary between fast and slow poloidal variation. Using trigonometric identities, we will incorporate the new poloidal coordinate by converting the flux surface specification from a 1-D Fourier series in $\theta$ to a 2-D Fourier series in $\theta$ and $z$. Equations \refEq{eq:fluxSurfaceSpec} and \refEq{eq:fluxSurfaceChangeSpec} become
\begin{align}
r_{0} \left( \theta, z \right) &= r_{\psi 0} \Bigg( 1 - \sum_{l = 0}^{\infty} \sum_{k = 0}^{m_{c} - 1} \frac{\Delta_{k + l m_{c}} - 1}{\Delta_{k + l m_{c}} + 1} \big[ \Cos{l \left( z + m_{c} \theta_{t m} \right)} \Cos{k \left( \theta + \theta_{t m} \right)} \nonumber \\
   -& \Sin{l \left( z + m_{c} \theta_{t m} \right)} \Sin{k \left( \theta + \theta_{t m} \right)} \big] \Bigg) \label{eq:fluxSurfaceSpecScaleSep} \\
\left. \frac{\partial r}{\partial r_{\psi}} \right|_{r_{\psi 0}, \theta, z} &= 1 - \sum_{l = 0}^{\infty} \sum_{k = 0}^{m_{c} - 1} \bigg[ \left( \frac{\Delta_{k + l m_{c}} - 1}{\Delta_{k + l m_{c}} + 1} + \frac{2 r_{\psi 0}}{\left( \Delta_{k + l m_{c}} + 1 \right)^{2}} \frac{d \Delta_{k + l m_{c}}}{d r_{\psi}} \right)  \label{eq:fluxSurfaceChangeSpecScaleSep} \\
   \times \big[& \Cos{l \left( z + m_{c} \theta_{t m} \right)} \Cos{k \left( \theta + \theta_{t m} \right)} - \Sin{l \left( z + m_{c} \theta_{t m} \right)} \Sin{k \left( \theta + \theta_{t m} \right)} \big] \bigg] \nonumber
\end{align}
respectively. The definition of $k \equiv m - l m_{c}$ is required to derive equations \refEq{eq:fluxSurfaceSpecScaleSep} and \refEq{eq:fluxSurfaceChangeSpecScaleSep} from equations \refEq{eq:fluxSurfaceSpec} and \refEq{eq:fluxSurfaceChangeSpec}. On the other hand, we define
\begin{align}
   l \equiv \left\lfloor \frac{m}{m_{c}} \right\rfloor \label{eq:lDef}
\end{align}
to reflect the physics of the scale separation as it formally divides the poloidal variation into fast and slow components. Here $\left\lfloor x \right\rfloor$ is the floor function that gives the integer value $K$ such that $K \leq x < K + 1$ for any real number $x$. This particular definition of $l$ means that variation at least as rapid as the $m_{c}$ Fourier mode is considered fast, while any slower variation is considered slow. However, the precise definition of $l$ does not matter as long as the fast and slow scales are sufficiently separated. For instance, in \cite{BallMirrorSymArg2016} we use numerical results to motivate the definition of $l \equiv \left\lfloor \left( m + 2 \right) / m_{c} \right\rfloor$.

The form of equations \refEq{eq:fluxSurfaceSpecScaleSep} and \refEq{eq:fluxSurfaceChangeSpecScaleSep} allow us to perform the separation of scales expansion needed for the tilting symmetry by expanding in $m_{c} \gg 1$. Introducing a poloidal translation of the fast variation through the substitution
\begin{align}
z \rightarrow z + z_{t} \label{eq:zDefTilted}
\end{align}
allows us to tilt only the fast flux surface variation, while keeping slowly-varying envelopes fixed. The tilting symmetry presented in \cite{BallMirrorSymArg2016} demonstrates that changing $z_{t}$ has only an exponentially small effect (in $m_{c} \gg 1$) on the turbulent fluxes. Equation \refEq{eq:zDefTilted} also permits us to formulate a mathematically rigorous definition of the envelope $r_{0, \text{env}} \left( \theta \right)$, a curve that is everywhere tangential to the family of curves generated by varying $z_{t}$ in $r_{0} \left( \theta, z + z_{t} \right)$. The envelope is given by
\begin{align}
   r_{0, \text{env}} \left( \theta \right) = r_{0} \left( \theta, z_{\text{env}} \left( \theta \right) \right) , \label{eq:env}
\end{align}
where $z_{\text{env}} \left( \theta \right)$ is a function calculated from
\begin{align}
   \left. \frac{\partial r_{0} \left( \theta, z_{\text{env}} \left( \theta \right) \right)}{\partial z_{\text{env}} \left( \theta \right)} \right|_{\theta} = 0 . \label{eq:envParameter}
\end{align}
In \ref{app:envCalc}, we will use these two equations to explicitly calculate $r_{0, \text{env}} \left( \theta \right)$ and $z_{\text{env}} \left( \theta \right)$ for the particular case of a flux surface with two shaping modes. These two equations also show that the geometry parameter $z_{t}$ does not enter into the calculation of the envelope. This is intuitive as tilting the fast flux surface variation by changing $z_{t}$ should not affect the envelope.

\subsection{Illustrative example}
\label{subsec:example}

Including only two shaping modes $\Delta_{6} = \Delta_{7} = 1.4$ in equation \refEq{eq:fluxSurfaceSpecScaleSep}, while setting the tilt angles to be $\theta_{t 6} = \theta_{t 7} = 0$ and $z_{t} = 0$, produces the cartoon flux surface shown in figure \ref{fig:envelopeShaping}(a). We see that the two poloidal shaping modes beat together to create an $m=1$ envelope. However, since the flux surface is exactly up-down symmetric, by the up-down symmetry argument \cite{PeetersMomTransSym2005, ParraUpDownSym2011, SugamaUpDownSym2011} we know it does not drive momentum transport.

To make use of the tilting symmetry, we will set $m_{c} = 6$ as a reasonable value for the characteristic mode number of the fast variation. By changing $z_{t} = \pi / 4$, we produce figure \ref{fig:envelopeShaping}(b), which is neither up-down symmetric, nor mirror symmetric. However, since we constructed it by tilting only the fast variation of an up-down symmetric flux surface, we know from the tilting symmetry argument that the momentum flux must be exponentially small in $m_{c} = 6 \gg 1$. The exponential scaling holds because the slow $m=1$ variation of the envelope was kept up-down symmetric, so it does not drive momentum transport.

This example shows that formally non-mirror symmetric configurations can still have exponentially small momentum flux if the slow variation is up-down symmetric and the fast variation is mirror symmetric. This becomes intuitive if we consider toroidicity as a second type of $m=1$ mode (in addition to the Shafranov shift). From this perspective up-down symmetry is just mirror symmetry with respect to the inherent, untilted mode from toroidicity. Hence, we can add any slow shaping mode as long as it is aligned with the mode from toroidicity, keeping the slow shaping mirror symmetric.

In order to tilt the entire flux surface and produce figure \ref{fig:envelopeShaping}(c), we must consider all flux surface variation ``fast.'' To do this we set $m_{c} = 1$ in equations \refEq{eq:fluxSurfaceSpecScaleSep} and \refEq{eq:fluxSurfaceChangeSpecScaleSep}. Taking $z_{t} = \pi / 4$ produces figure \ref{fig:envelopeShaping}(c), which is a mirror symmetric flux surface. However, since we rotated the $m=1$ variation of the envelope, we have considered $m=1$ variation to be fast. This means that (though the tilting symmetry technically still holds) the difference in the fluxes produced by figures \ref{fig:envelopeShaping}(a) and \ref{fig:envelopeShaping}(c) is exponentially small in an expansion in $m_{c} = 1 \gg 1$ (which is not particularly meaningful).

If a flux surface is mirror symmetric it produces momentum transport that is exponentially small in the Fourier mode number of the poloidal variation. Hence, the tilting symmetry argument creates a distinction between mirror and non-mirror symmetric flux surface shapes. However, this example illustrates that the validity of this distinction becomes doubtful in certain cases, specifically geometries with low shaping effects or a slowly-varying envelope created by the beating of several high mode number effects.

\subsection{Envelopes}
\label{subsec:envelopes}

In order to understand the effect of an up-down asymmetric, slowly-varying envelope, we will use the analytic calculation presented in section 2.4 of \cite{BallMomFluxScaling2016}. The calculation considers a flux surface with up-down symmetric slow shaping modes and completely general fast shaping modes with a characteristic magnitude of $\Delta_{m} - 1$. Note that it does not assume that slow variation created by fast shaping modes (i.e. an envelope) is necessarily up-down symmetric. Using this geometry, the scaling of the symmetry-breaking in the eight geometric coefficients is derived in the $m_{c} \gg 1$ limit. The calculation is summarized by table 1 of \cite{BallMomFluxScaling2016} and demonstrates that \emph{in general} the up-down symmetry of the geometric coefficients is broken to $\Order{m_{c}^{3} \left( \Delta_{m} - 1 \right)^{2}}$, a polynomial order in $m_{c} \gg 1$. It is then proven that this leads to momentum flux that is also $\Order{m_{c}^{3} \left( \Delta_{m} - 1 \right)^{2}}$ (i.e. polynomially small). The calculation also notes that $\Delta_{m} - 1 \sim m_{c}^{-2}$ is a reasonable and physical scaling, which leads to $\Order{m_{c}^{-1}}$ momentum flux.

This result can still be consistent with the exponential scalings proven using the tilting symmetry, so long as all the polynomial symmetry-breaking terms cancel for the particular flux surface shape. In other words, we expect a polynomial scaling in general, but given a specific geometry all the polynomial symmetry-breaking terms may cancel (leading to an exponential scaling). Looking at the calculation in section 2.3 of \cite{BallMomFluxScaling2016} we can argue that such a cancellation does occur in geometries that follow the tilting symmetry (e.g. figure \ref{fig:envelopeShaping}(b)).

Section 2.3 of \cite{BallMomFluxScaling2016} performs a similar calculation to section 2.4 of \cite{BallMomFluxScaling2016}, but to lowest order in aspect ratio with a specific geometry. The geometry only includes two fast shaping modes at $m$ and $n$, both with a magnitude that scales as $\Delta_{m} - 1 \sim \Delta_{n} - 1 \sim m_{c}^{-2}$. This concrete geometry specification is general enough to create envelopes, but is simple enough to allow the calculation of the eight geometric coefficients (as opposed to just their scalings with $m_{c} \gg 1$). The coefficient $\vec{v}_{d s} \cdot \Nabla \alpha$ is explicitly derived as an example, while the results for all coefficients are stated in Appendix D of \cite{BallMomFluxScaling2016}. To lowest order in $m_{c} \gg 1$, the calculation finds the usual result for circular flux surfaces,
\begin{align}
   \left( \vec{v}_{d s} \cdot \Nabla \alpha \right)_{0} &= \frac{B_{0}}{R_{0} \Omega_{s}} \left( \frac{d \psi}{d r_{\psi}} \right)^{-1} \left( \Cos{\theta} + \hat{s}' \theta \Sin{\theta} \right) ,
\end{align}
where $B_{0}$ is the value of the toroidal magnetic field at $R = R_{0}$, $\Omega_{s}$ is the gyrofrequency, and $\hat{s}'$ (defined by equation \refEq{eq:shiftedShatDef}) is related to the magnetic shear. To next order, reference \cite{BallMomFluxScaling2016} calculates a complicated expression with the form of
\begin{align}
   \left( \vec{v}_{d s} \cdot \Nabla \alpha \right)_{1} &= D_{1} \theta \Sin{\theta} + \left( D_{2} \Sin{\theta} + D_{3} \theta \Cos{\theta} \right) \left( D_{4} \Sin{z_{m s}} + D_{5} \Sin{z_{n s}} \right) \nonumber \\
   &+ \left( D_{6} \Cos{\theta} + D_{7} \theta \Sin{\theta} \right) \left( D_{8} \Cos{z_{m s}} + D_{9} \Cos{z_{n s}} \right) + D_{10} \Sin{\theta} \label{eq:nonMirrorAlphaDrift} \\
   &\times \left[ \Sin{\left( n - m \right) \theta} \Cos{m \left( \theta_{t m} - \theta_{t n} \right)} - \Cos{\left( n - m \right) \theta} \Sin{m \left( \theta_{t m} - \theta_{t n} \right)} \right] , \nonumber
\end{align}
where
\begin{align}
  z_{m s} &\equiv m \left( \theta + \theta_{t m} \right) \label{eq:zmsDef} \\
  z_{n s} &\equiv n \left( \theta + \theta_{t n} \right) \label{eq:znsDef}
\end{align}
and $D_{i}$ are constants (the full expression is given in Appendix D of \cite{BallMomFluxScaling2016}). We see that, even after averaging over $z$, the last term remains. This term has the coefficient
\begin{align}
   D_{10} = \frac{r_{\psi 0}}{\left( n - m \right)} \left( m^{2} \left( \Delta_{m} - 1 \right) \frac{d \Delta_{n}}{d r_{\psi}} + n^{2} \left( \Delta_{n} - 1 \right) \frac{d \Delta_{m}}{d r_{\psi}} \right) , \label{eq:symBreakCoeff}
\end{align}
indicating that it results from the beating between the two shaping effects. This term breaks the up-down symmetry of the gyrokinetic equation to a polynomial order in $m_{c} \gg 1$ and generates momentum transport that is $\Order{m_{c}^{-1}}$.
 
However, in order to understand the effect of envelopes, we will use a less manipulated form of the geometric coefficients (given in \ref{app:nonMirrorGeoCoeffs}) to make an important clarification. Here we give one coefficient,
\begin{align}
   \left( \vec{v}_{d s} \cdot \Nabla \alpha \right)_{1} &= D_{1} \theta \Sin{\theta} + \left( D_{2} \Sin{\theta} + D_{3} \theta \Cos{\theta} \right) \left( D_{4} \Sin{z_{m s}} + D_{5} \Sin{z_{n s}} \right) \nonumber \\
   &+ \left( D_{6} \Cos{\theta} + D_{7} \theta \Sin{\theta} \right) \left( D_{8} \Cos{z_{m s}} + D_{9} \Cos{z_{n s}} \right) + D_{10} \Sin{\theta} \label{eq:nonMirrorAlphaDriftReworked} \\
   &\times \left[ \Sin{\left( n - m \right) \theta} \Cos{m \theta_{t m} - n \theta_{t n}} - \Cos{\left( n - m \right) \theta} \Sin{m \theta_{t m} - n \theta_{t n}} \right] , \nonumber
\end{align}
as an example. The only discrepancies between equations \refEq{eq:nonMirrorAlphaDrift} and \refEq{eq:nonMirrorAlphaDriftReworked} are in the arguments of the trigonometric functions of the symmetry-break terms (i.e. the last two terms). The difference between the arguments of $m \left( \theta_{t m} - \theta_{t n} \right)$ and $m \theta_{t m} - n \theta_{t n}$ is $\left( m - n \right) \theta_{t n}$. Since $\theta_{t n} \in \left[ 0 , 2 \pi / n \right)$ and $m \sim n \sim m_{c}$, this term is small by one order in $m_{c} \gg 1$. Hence, deriving equation \refEq{eq:nonMirrorAlphaDrift} from equation \refEq{eq:nonMirrorAlphaDriftReworked} only introduces an $\Order{m_{c}^{-2}}$ error. However, if we want to determine when the symmetry-breaking terms cancel to \emph{all} polynomial orders an $\Order{m_{c}^{-2}}$ term matters, so we must use the form of equation \refEq{eq:nonMirrorAlphaDriftReworked}.

From equations \refEq{eq:symBreakCoeff} and \refEq{eq:nonMirrorAlphaDriftReworked} we see that the symmetry-breaking terms cancel if $n = m$, $\Delta_{m} = 1$, $\Delta_{n} = 1$, or $\Sin{m \theta_{t m} - n \theta_{t n}} = 0$. The first three conditions imply that a slowly-varying envelope does not exist, but the fourth condition is different. It indicates that if
\begin{align}
   \theta_{t n} = \frac{m}{n} \theta_{t m} + K_{1} \frac{\pi}{n} \label{eq:twoModeEnvelopeCond}
\end{align}
for some integer $K_{1}$, then the momentum transport is no longer polynomial and could be exponential. As demonstrated by the derivation in \ref{app:envCalc}, a flux surface with a slowly-varying envelope fulfills this condition if and only if the envelope is up-down symmetric (e.g. figure \ref{fig:envelopeShaping}(a,b)). Hence, if a slowly-varying envelope is present and it is not up-down symmetric, then the momentum flux is only polynomially small.

In summary, we generally expect the momentum transport driven by fast flux surface shaping (i.e. on the order of $m_{c}$) to be polynomially small in $m_{c} \gg 1$. However, if a flux surface has mirror symmetric fast shaping modes, then the momentum transport is further restricted to be exponentially small in $m_{c} \gg 1$, unless the fast variation creates an up-down asymmetric slowly-varying envelope.

\section{Numerical gyrokinetic analysis}
\label{sec:numerics}

In this section we will present nonlinear gyrokinetic simulations from the local $\delta f$ code GS2 \cite{DorlandETGturb2000} to test the analytic conclusions of section \ref{sec:analytics}. We will use the two sets of simulations shown in section 3 of \cite{BallMomFluxScaling2016}, but must add a third set in order to properly test the scalings.

\begin{figure}
 \centering

 \includegraphics[width=0.185\textwidth]{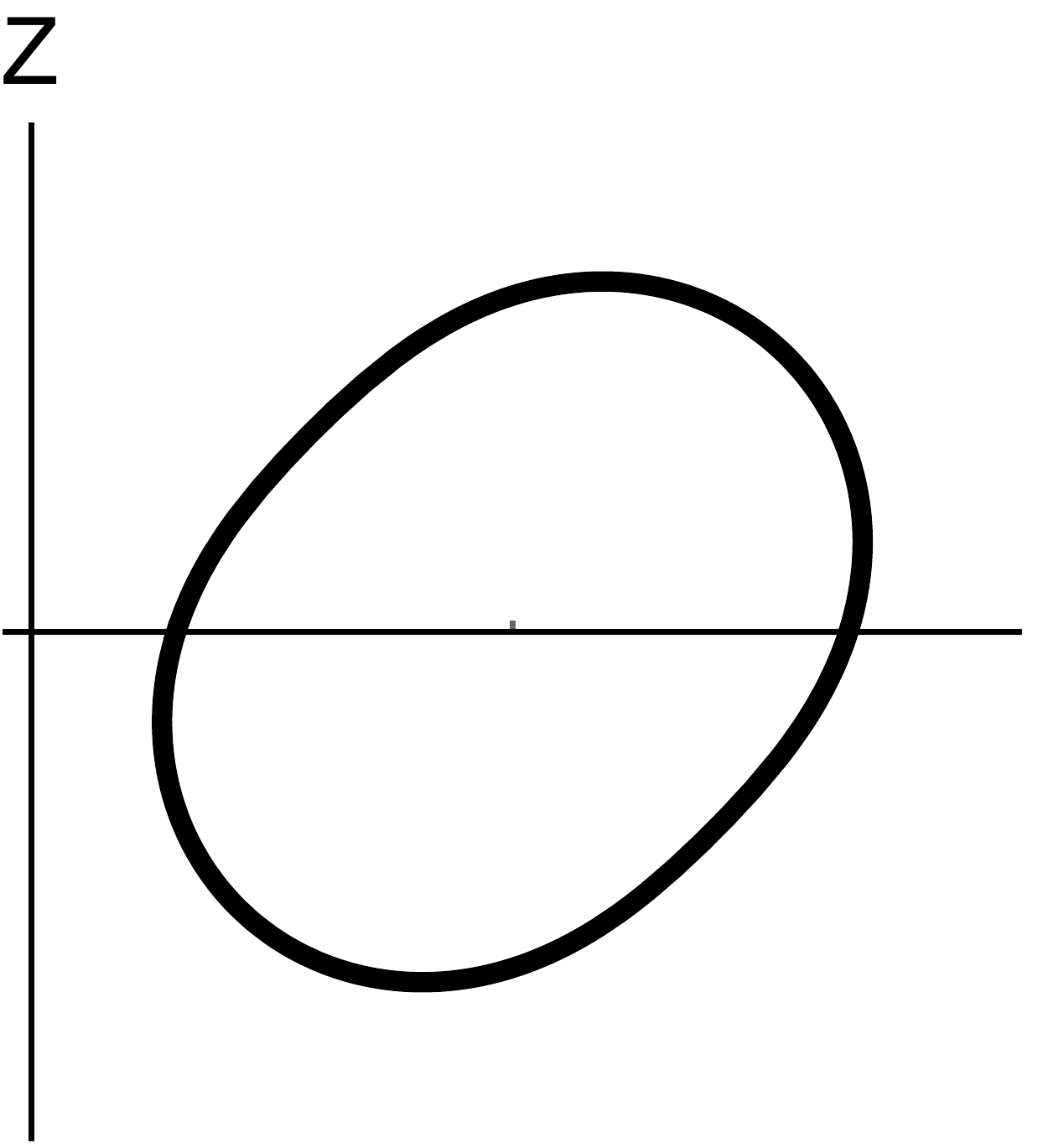}
 \includegraphics[width=0.185\textwidth]{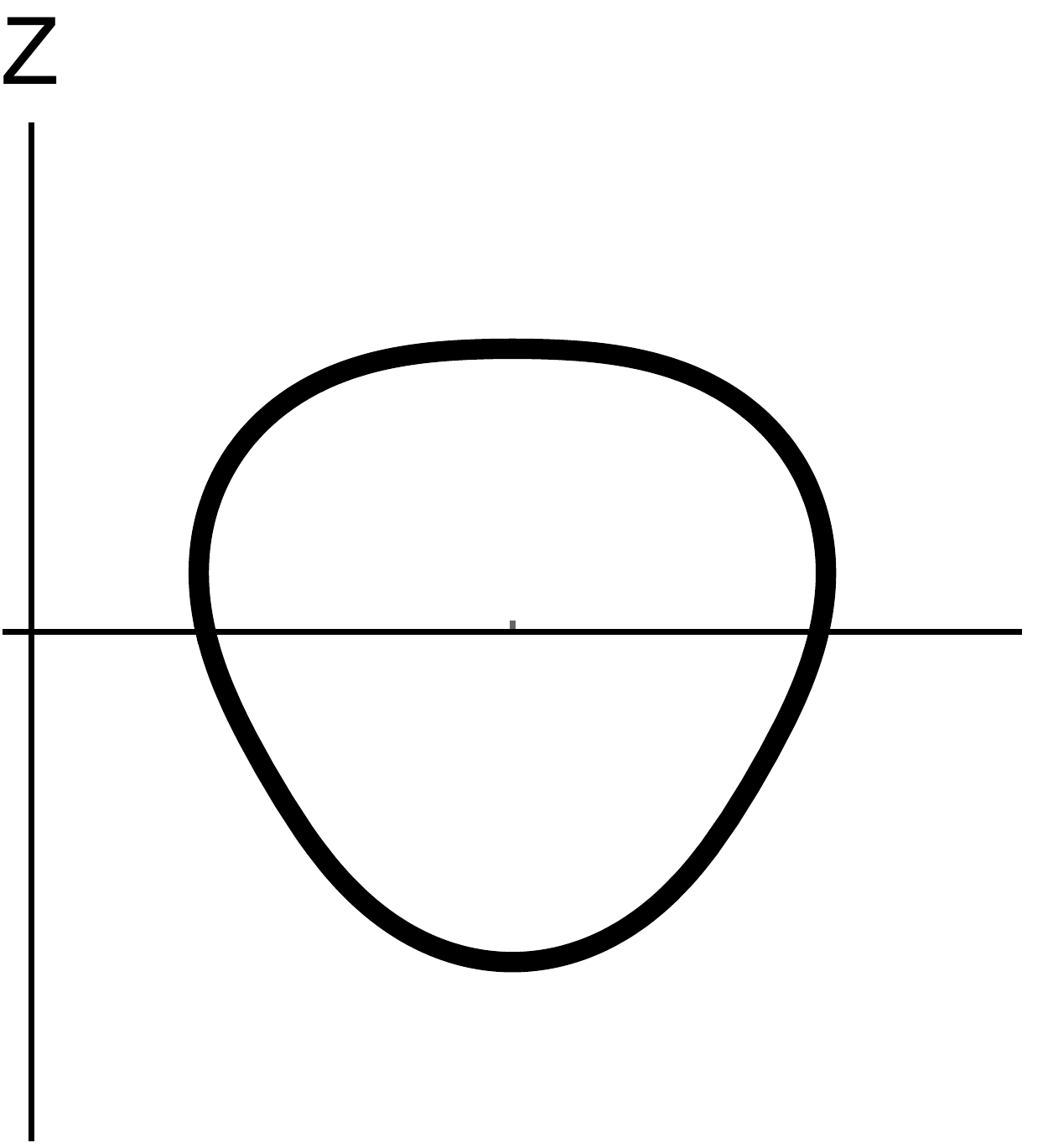}
 \includegraphics[width=0.185\textwidth]{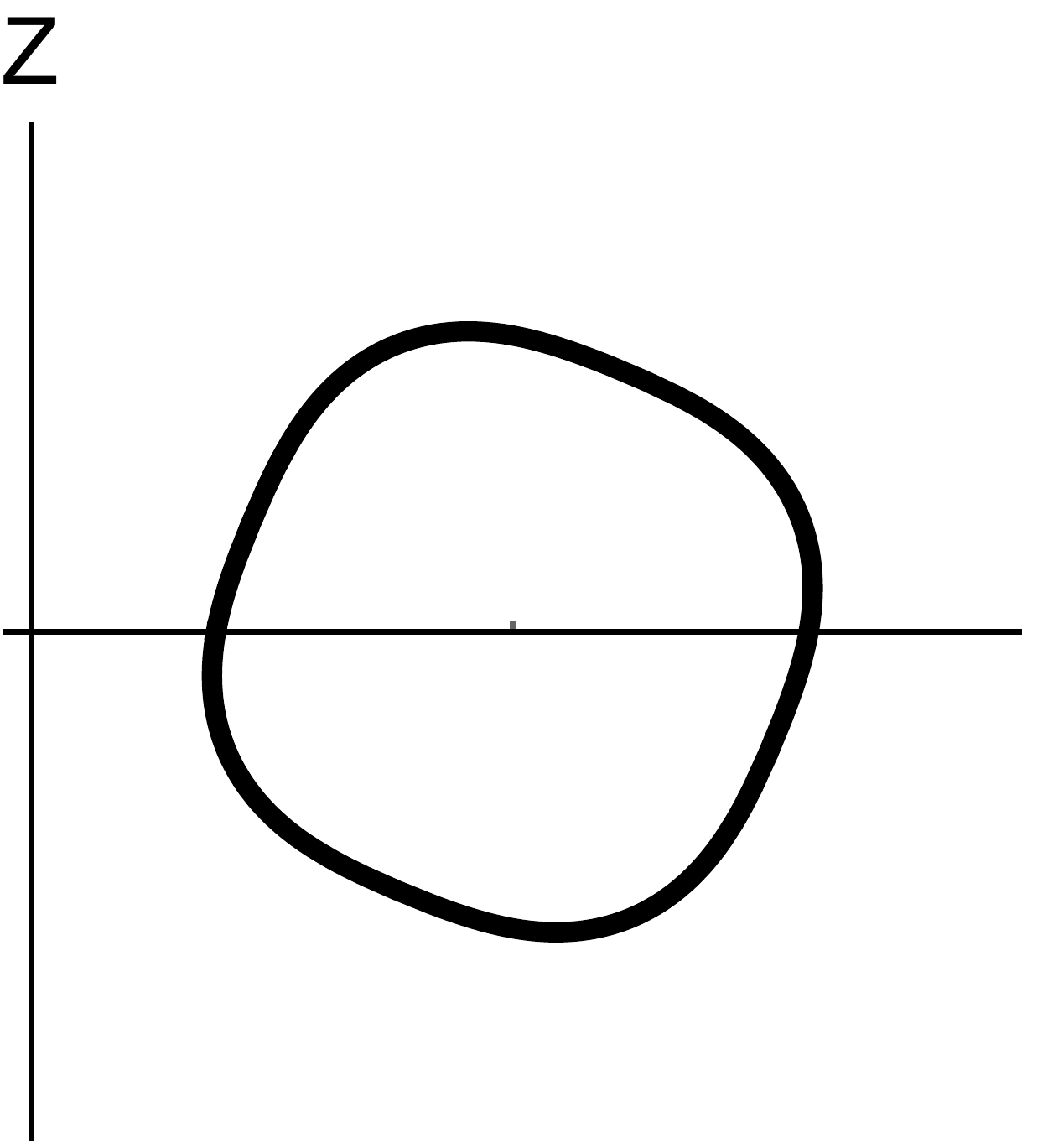}
 \includegraphics[width=0.185\textwidth]{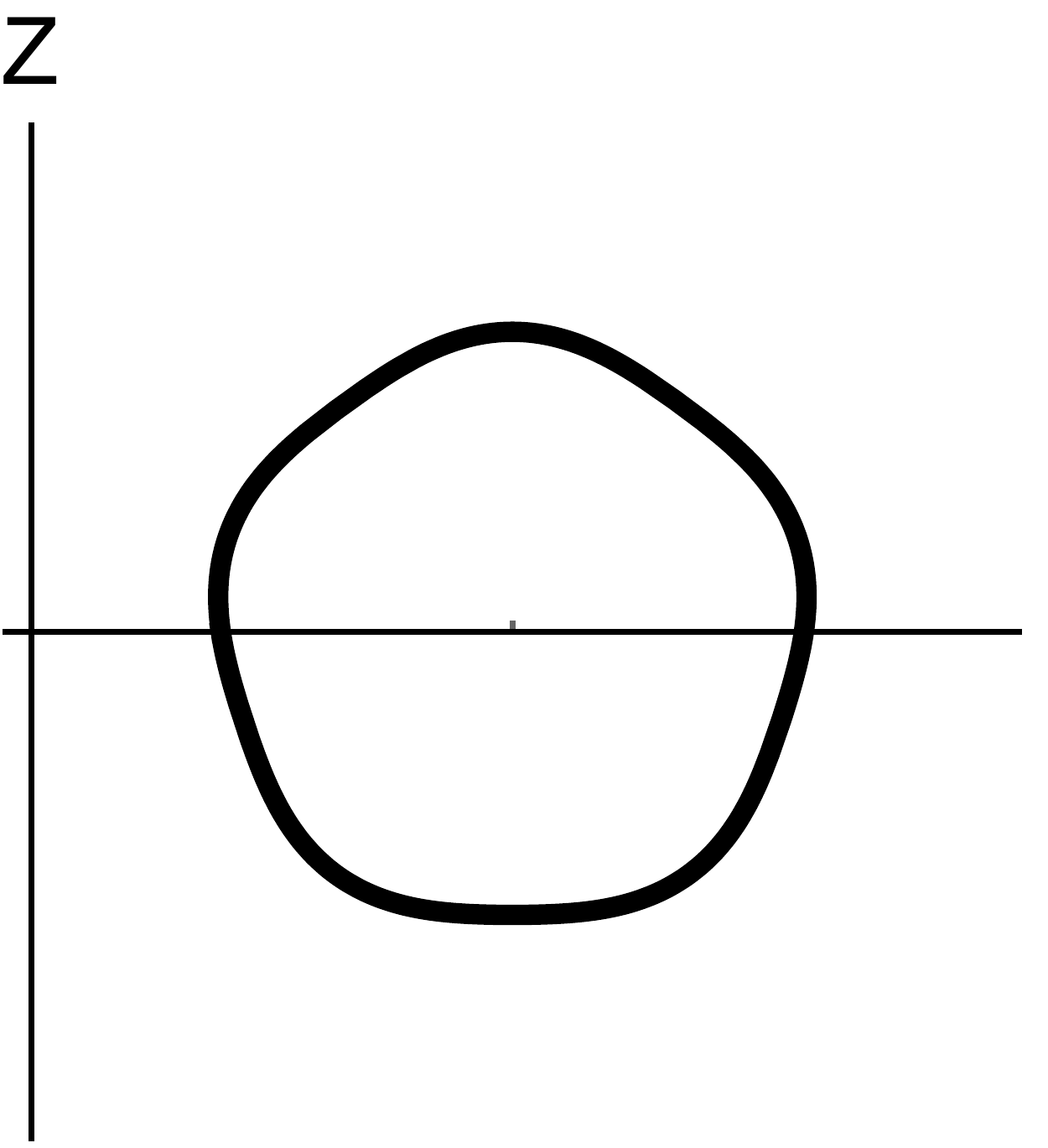}
 \includegraphics[width=0.185\textwidth]{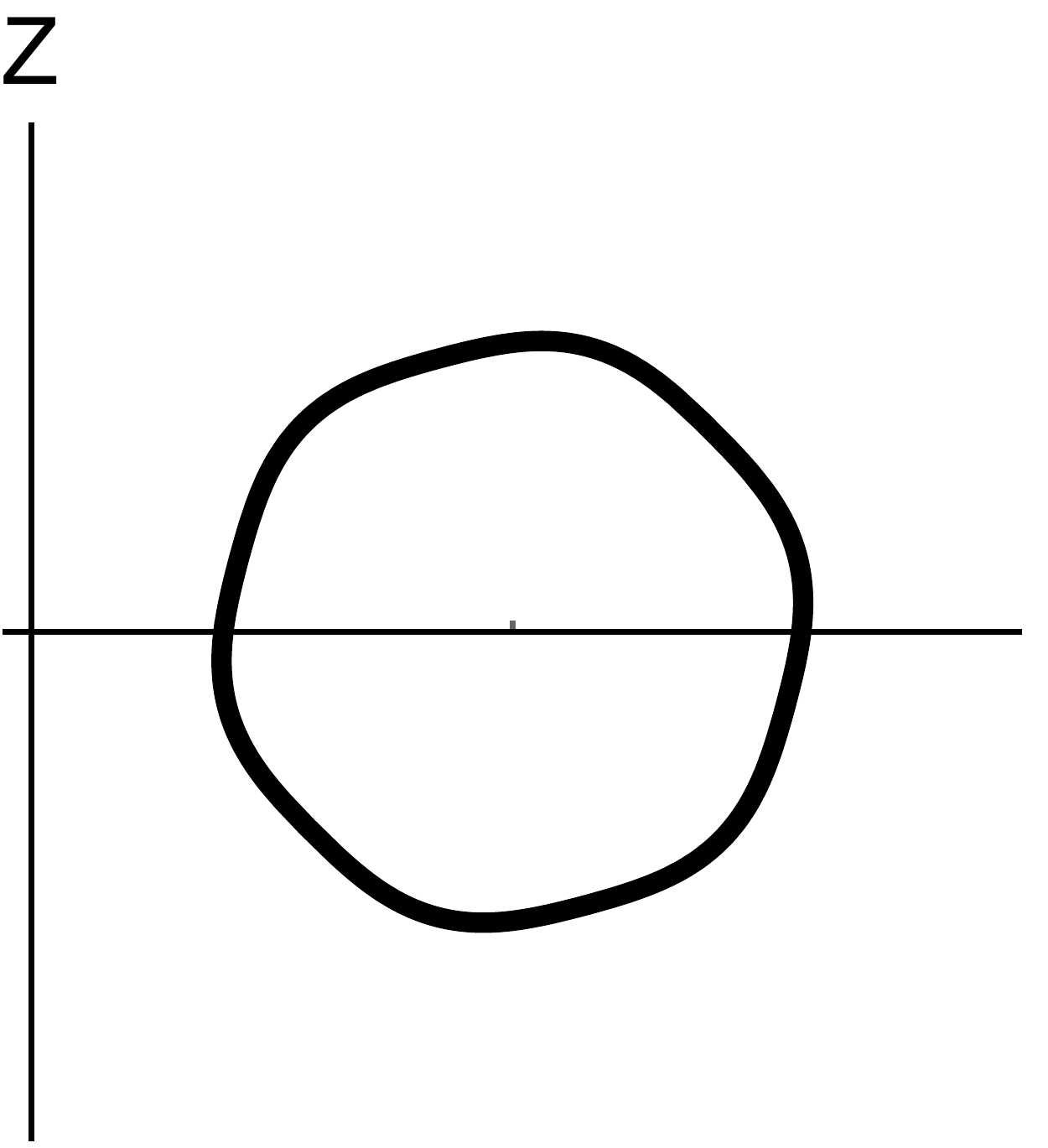}
 \includegraphics[width=0.0113\textwidth]{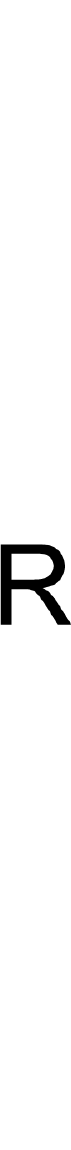}
 
 
 \includegraphics[width=0.185\textwidth]{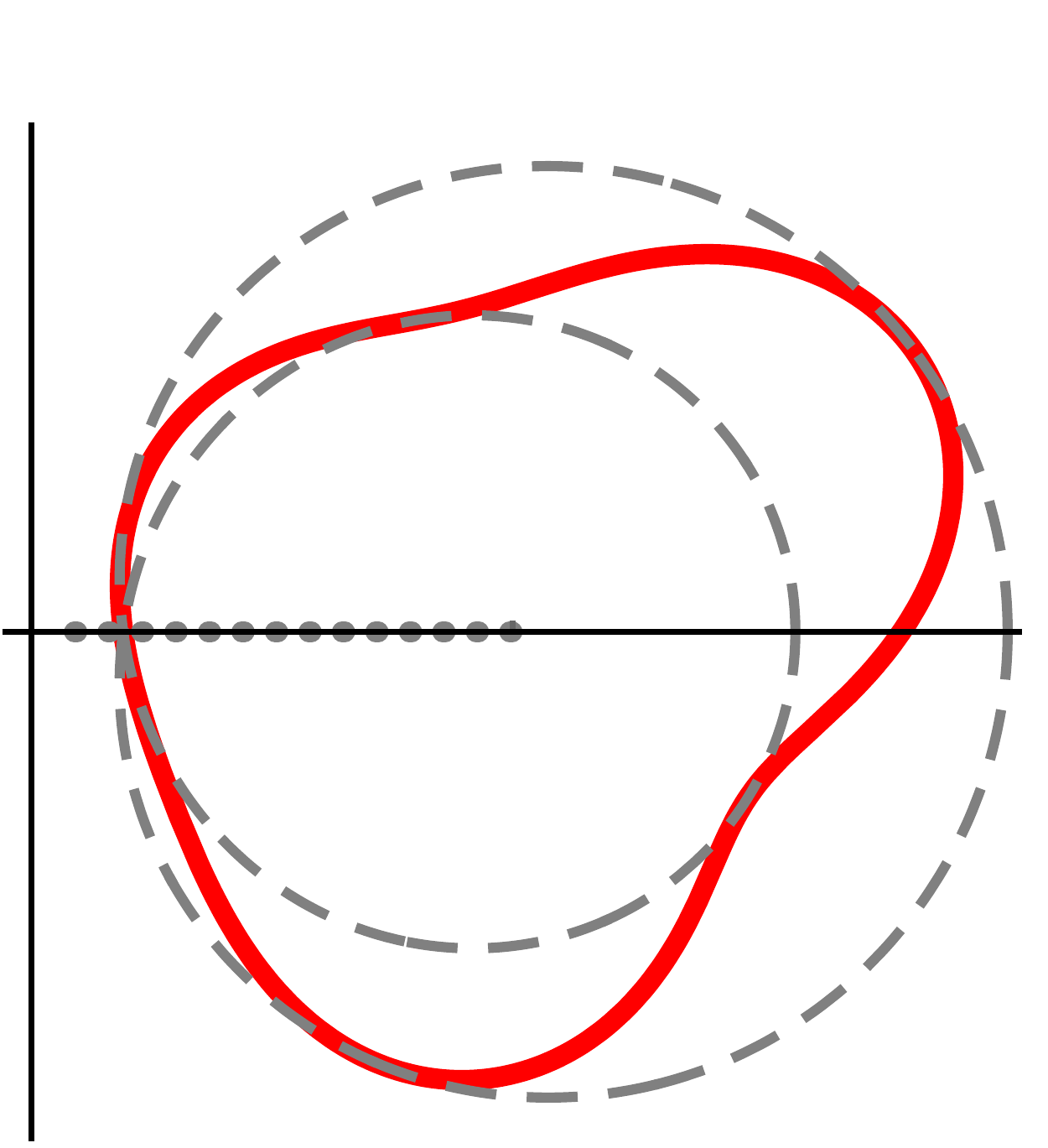}
 \includegraphics[width=0.185\textwidth]{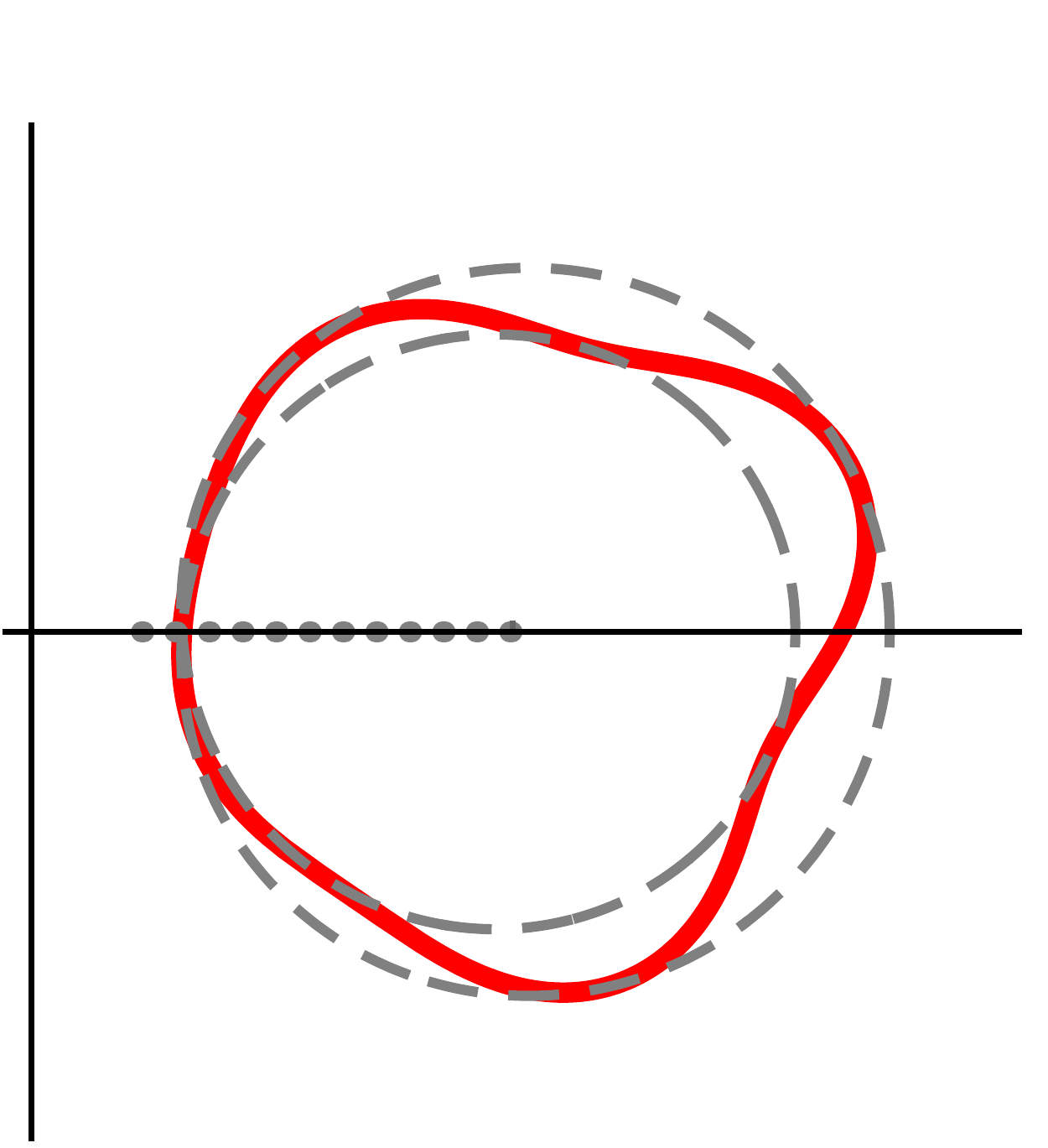}
 \includegraphics[width=0.185\textwidth]{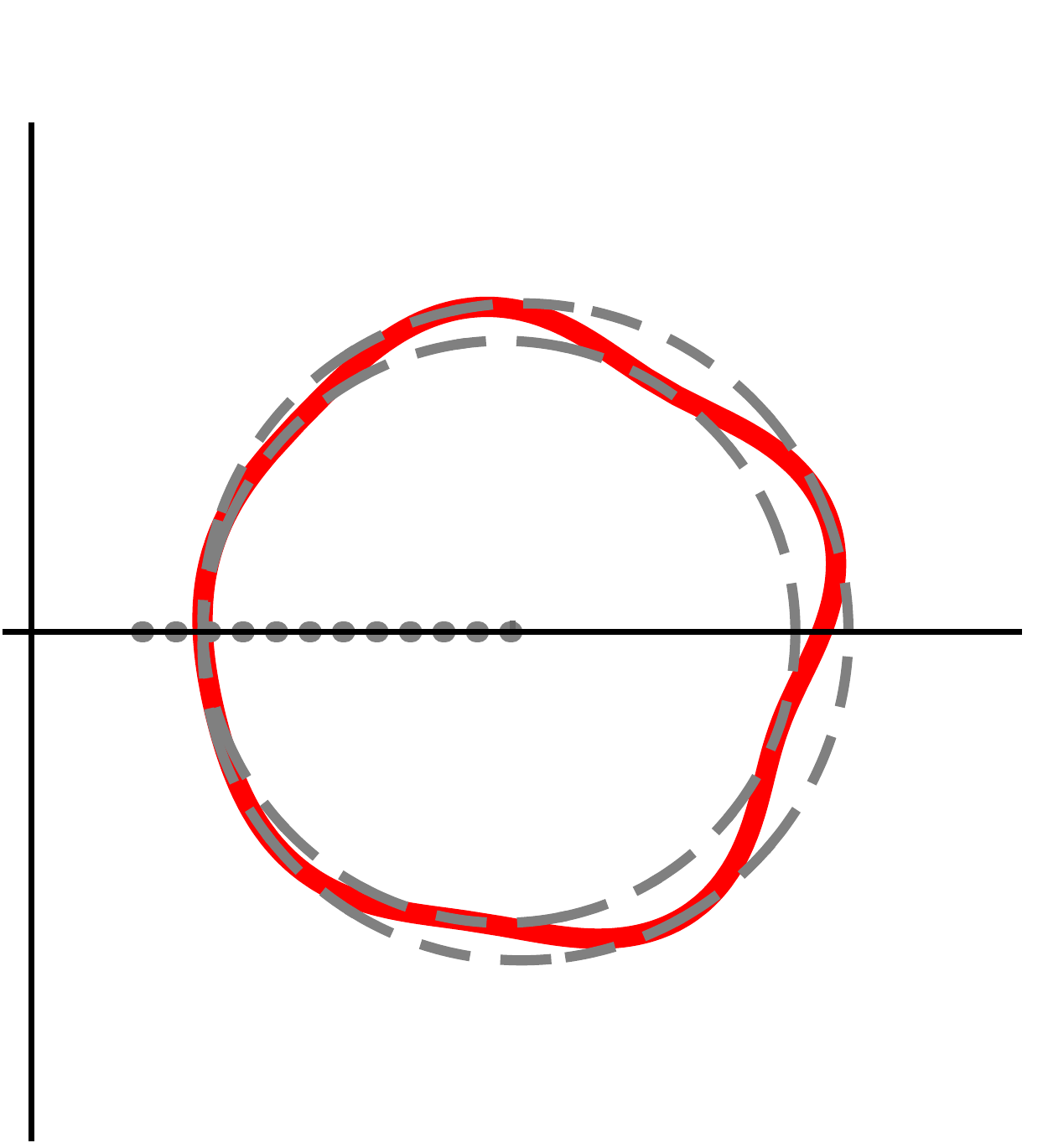}
 \includegraphics[width=0.185\textwidth]{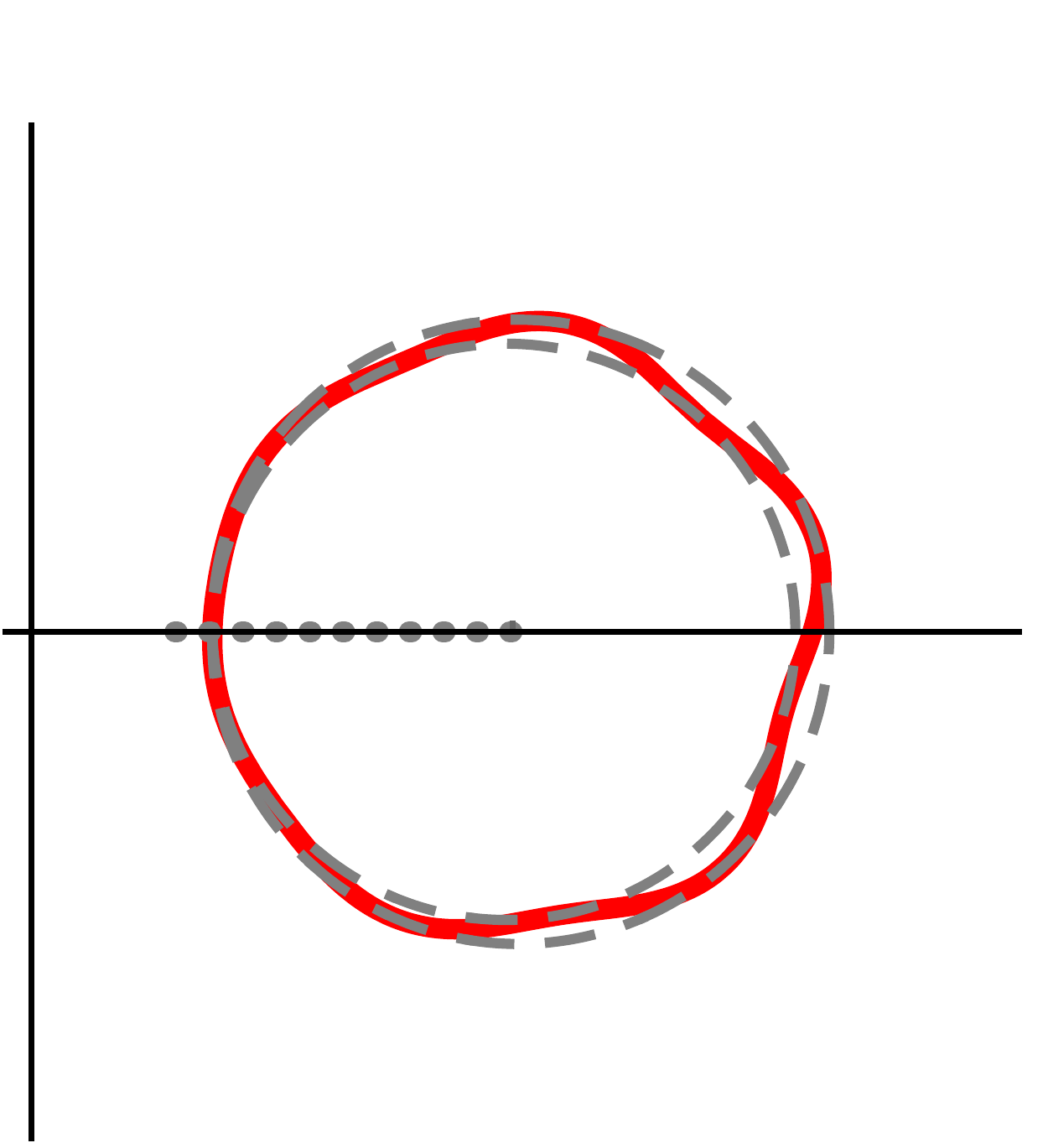}
 \includegraphics[width=0.185\textwidth]{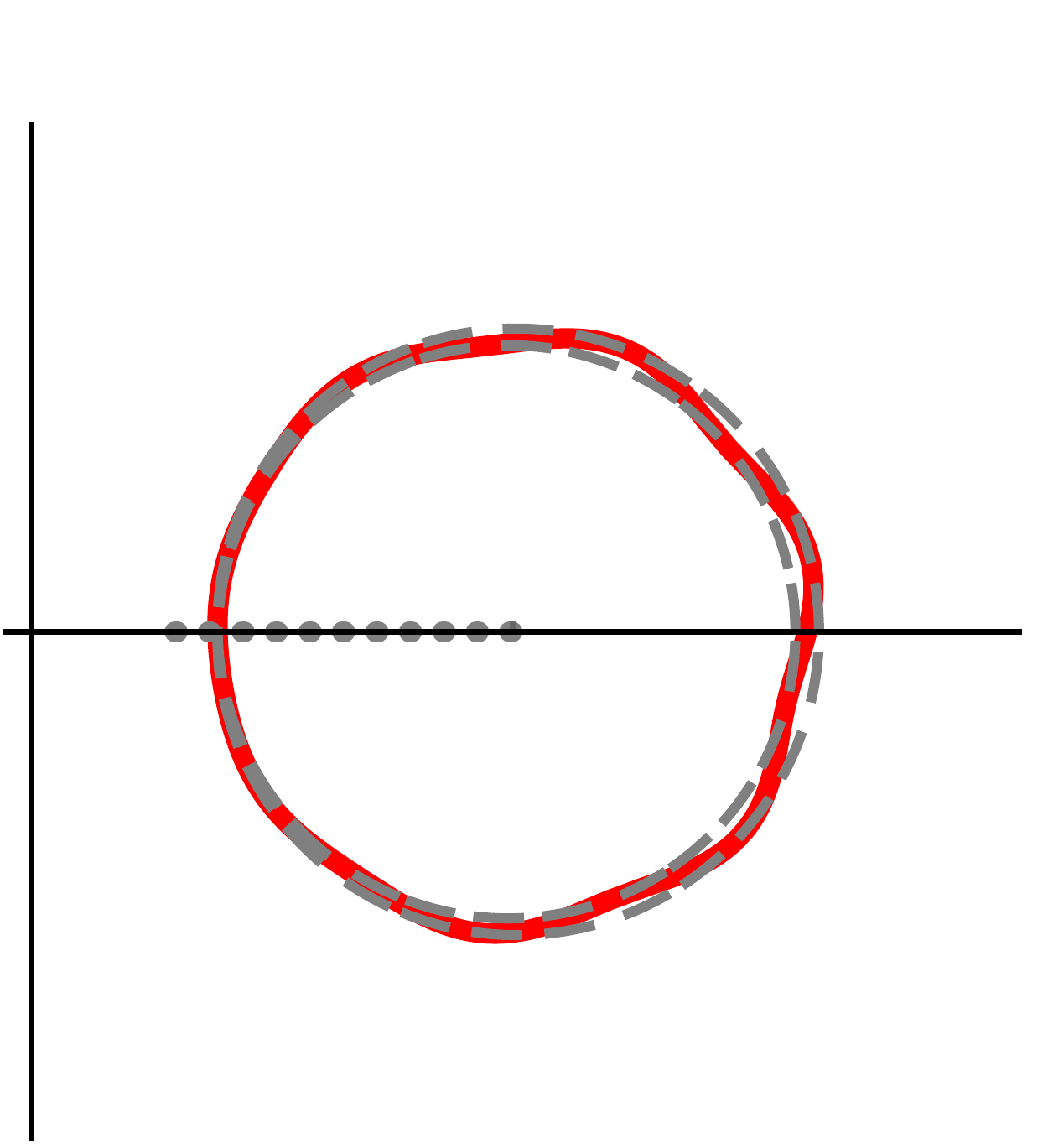}
 \includegraphics[width=0.0113\textwidth]{figs/xAxisLabelR.pdf}
 
 \includegraphics[width=0.185\textwidth]{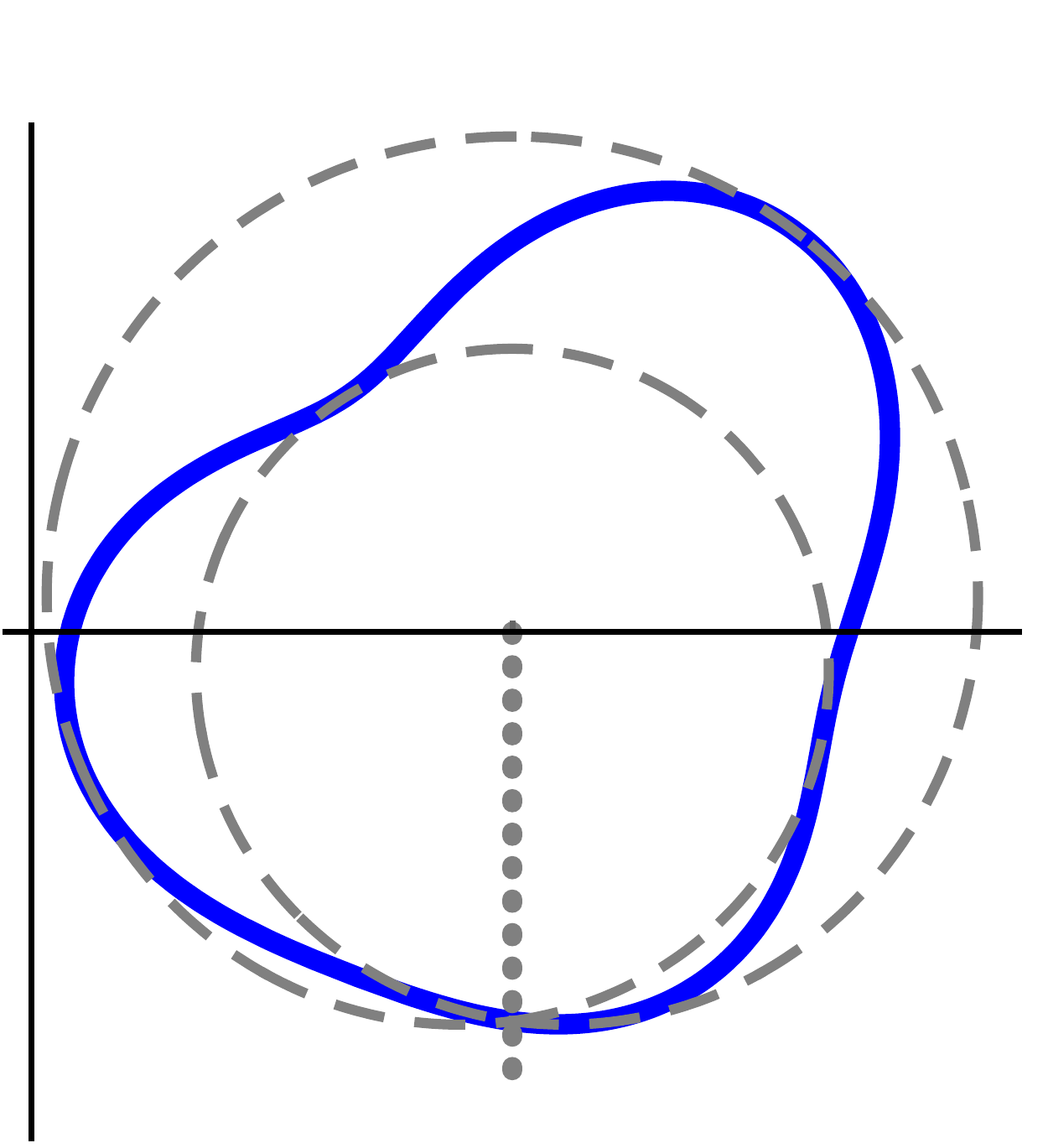}
 \includegraphics[width=0.185\textwidth]{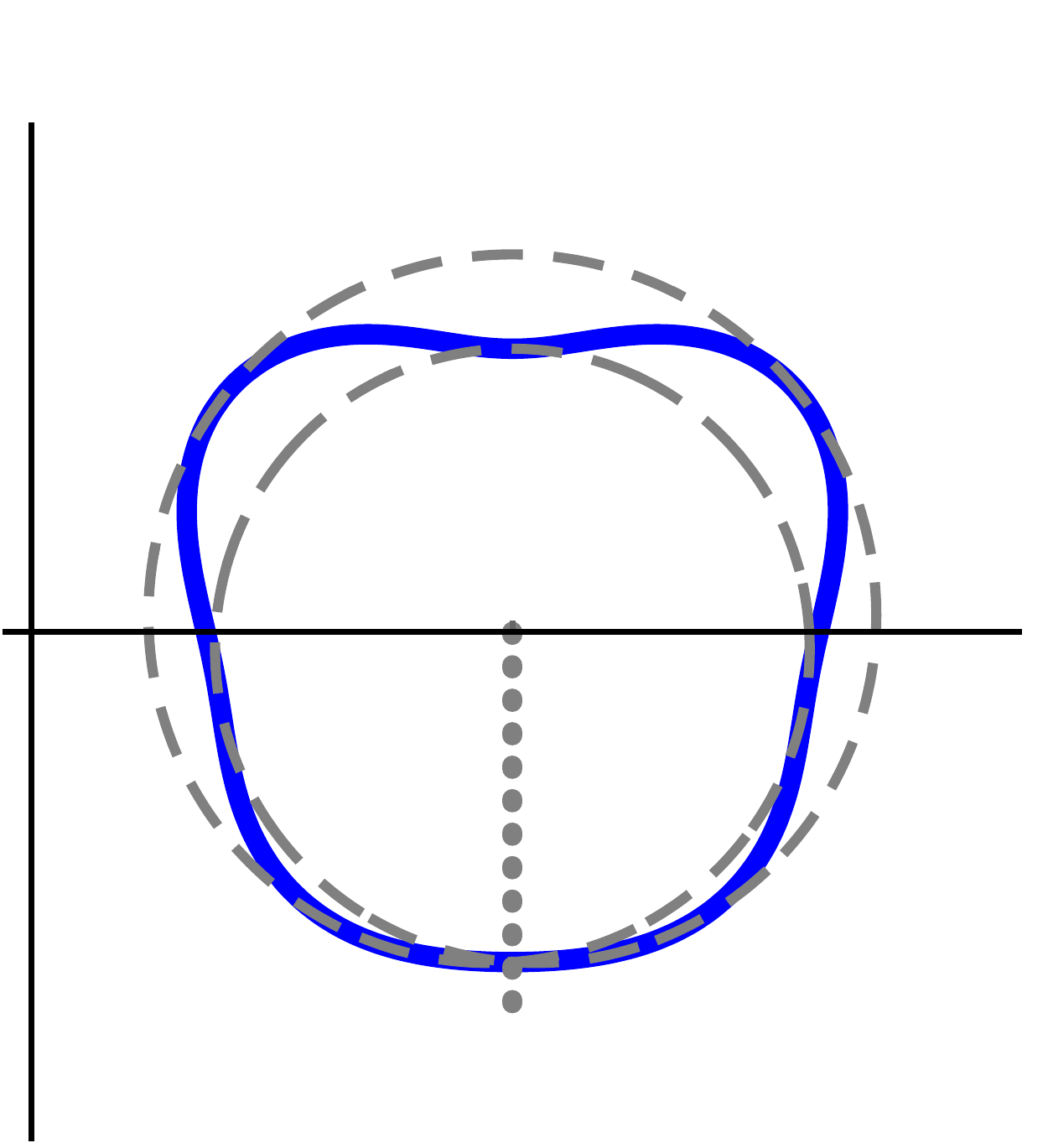}
 \includegraphics[width=0.185\textwidth]{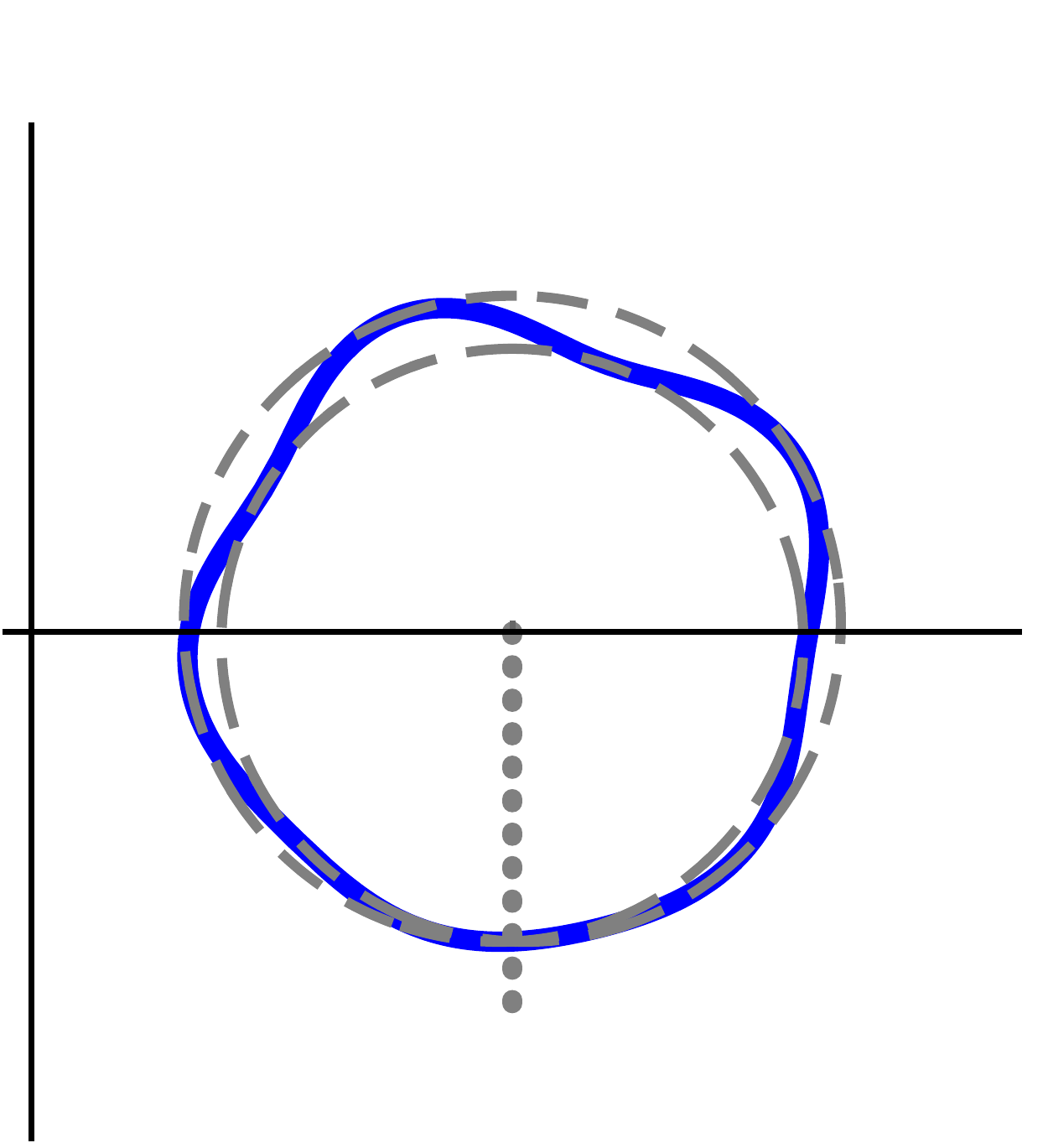}
 \includegraphics[width=0.185\textwidth]{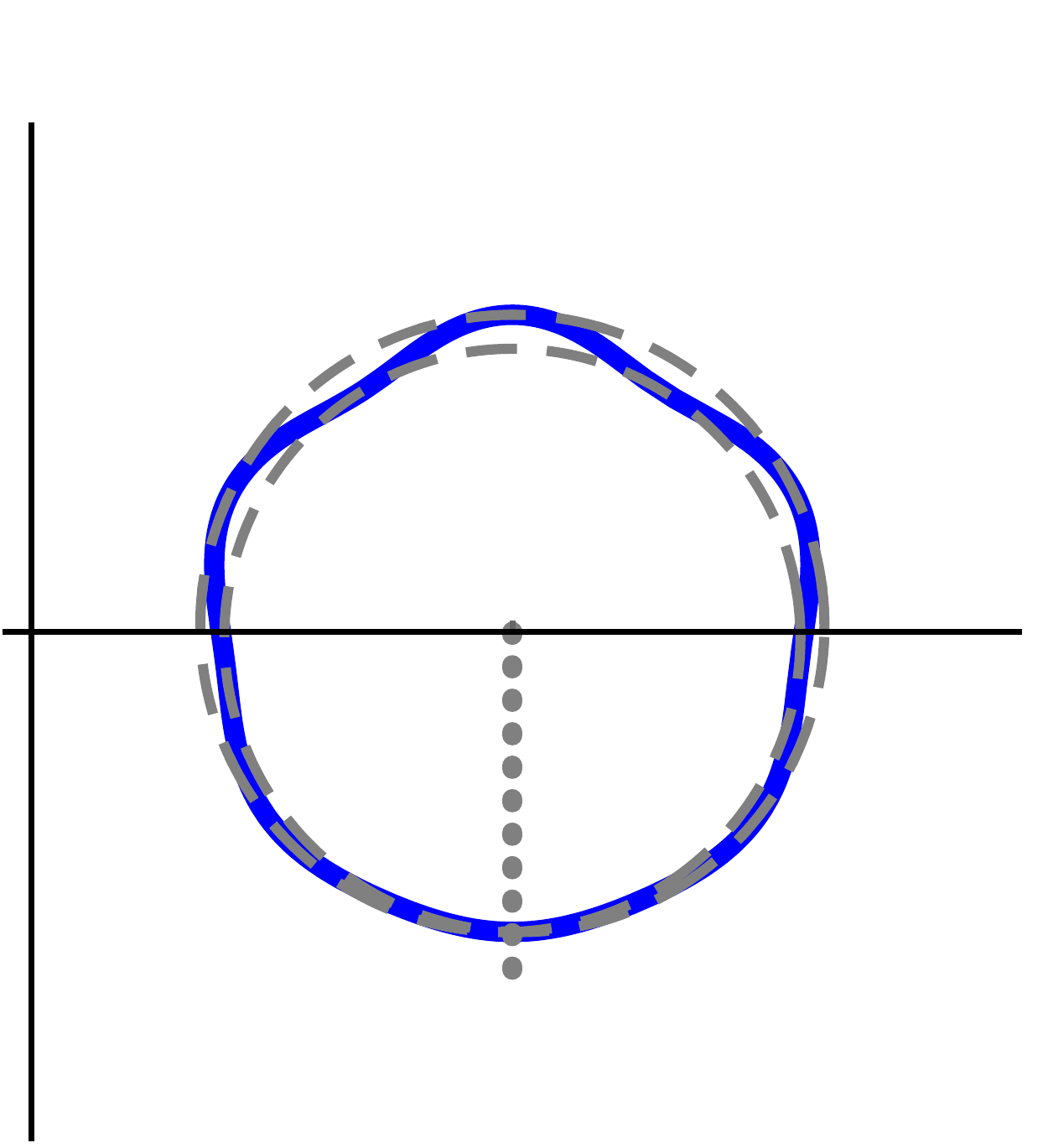}
 \includegraphics[width=0.185\textwidth]{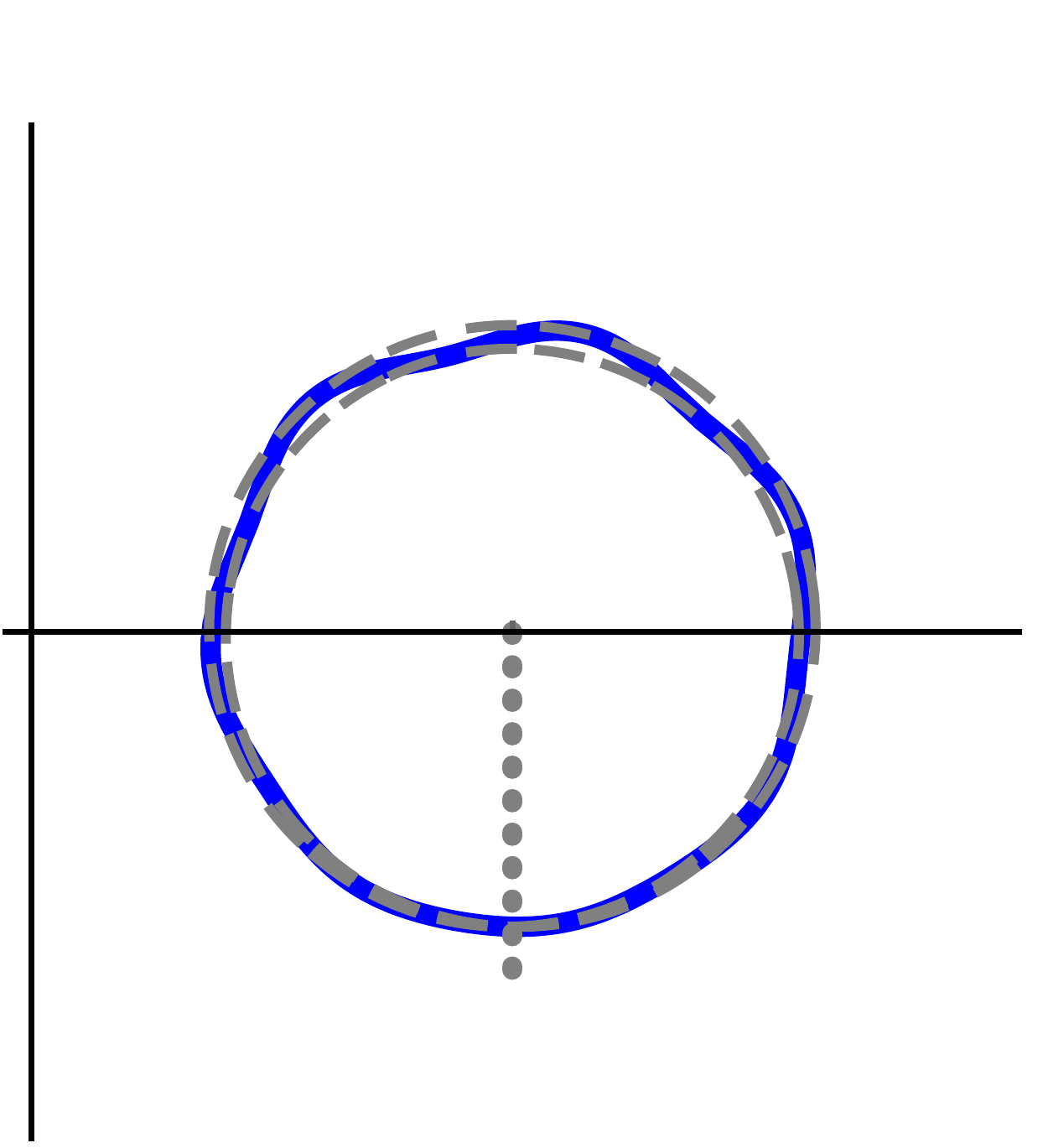}
 \includegraphics[width=0.0113\textwidth]{figs/xAxisLabelR.pdf}

 \caption{The $m_{c}=2$ through $m_{c}=6$ flux surface geometries in the mirror symmetric (top row, black), non-mirror symmetric with up-down symmetric envelopes (middle row, red) and non-mirror symmetric with up-down asymmetric envelopes (bottom row, blue) scans, where the envelope (gray, dashed) and the envelope tilt angle (gray, dotted) are shown if present.}
 \label{fig:simGeo}
\end{figure}

We will specify the three sets of flux surface shapes using equations \refEq{eq:fluxSurfaceSpec} and \refEq{eq:fluxSurfaceChangeSpec}. Each set will include several simulations at different values of $m_{c}$ in order to directly check the scalings of turbulent transport with $m_{c} \gg 1$. The geometries for the three scans are shown in figure \ref{fig:simGeo}. They include mirror symmetric flux surfaces, non-mirror symmetric flux surfaces with an up-down symmetric envelope, and non-mirror symmetric flux surfaces with an up-down asymmetric envelope.

All simulations include at least one shaping mode $m_{c}$ with a magnitude of $m_{c}^{2} \left( \Delta_{m} - 1 \right) = 1.5$. We have chosen to scale $\Delta_{m} - 1 \sim m_{c}^{-2}$ because exceeding this necessarily leads to flux surfaces with convex regions (see section 2.3 of \cite{BallMomFluxScaling2016} for more details). The tilt angle of this mode is set to be $\theta_{t m} = \pi / \left( 2 m \right)$ because it is halfway between neighboring up-down symmetric configurations (at $\theta_{t m} = 0$ and $\theta_{t m} = \pi / m$). The non-mirror symmetric geometries use a second mode $n = m_{c} + 1$ with a strength of $n^{2} \left( \Delta_{n} - 1 \right) = 1.5$. For geometries with up-down symmetric envelopes, we set $\theta_{t n} = \pi / \left( 2 n \right)$ to be consistent with equation \refEq{eq:twoModeEnvelopeCond}. For the geometries with up-down asymmetric envelopes we set $\theta_{t n} = 0$ because it is halfway between neighboring configurations with up-down symmetric envelopes (at $\theta_{t n} = \pi / \left( 2 n \right)$ and $\theta_{t n} = - \pi / \left( 2 n \right)$).

We note that \cite{BallMomFluxScaling2016} only included a mirror symmetric and one non-mirror symmetric scan. However, the non-mirror symmetric scan had tilt angles of $\theta_{t m} = \pi / \left( 2 m \right)$ and $\theta_{t n} = \theta_{t m} - \pi / \left( 2 m n \right)$ in order to be halfway between neighboring mirror symmetric configurations. Unfortunately, for the specific case of $n = m_{c} + 1$, these angles happen to create an exactly up-down symmetric envelope. That means that the caption of figure 4 is incorrect. We expect an exponential scaling for the specific geometries used in the non-mirror symmetric scan, not a $m_{c}^{-1}$ scaling. Hence, here we have added a second non-mirror symmetric scan, which has an up-down asymmetric envelope and therefore a $m_{c}^{-1}$ scaling.

All simulations in this work are electrostatic and collisionless with deuterium ions and kinetic electrons. They were run with a resolution of at least 48 poloidal grid points, 127 radial wavenumber grid points, 22 poloidal wavenumber grid points, 12 energy grid points, and 10 untrapped pitch angle grid points. At a given value of $m_{c}$, we ensured that the two non-mirror symmetric simulations had identical resolutions and were run to similar in-simulation times. Unless otherwise specified all simulations use Cyclone base case parameters \cite{DimitsCycloneBaseCase2000}: the minor radius $r_{\psi 0} / a= 0.54$, the major radial location of the flux surface of interest $R_{0} / a = 3$, the safety factor $q = 1.4$, the magnetic shear $\hat{s} = 0.8$, the temperature gradient $a / L_{T s} = 2.3$, and the density gradient $a / L_{n s} = 0.733$. The effect of the pressure gradient on the magnetic equilibrium is ignored (i.e. $d p / d \psi = 0$). For certain non-mirror symmetric geometries, the turbulence was stabilized at the Cyclone base case temperature gradient. Therefore, all non-mirror symmetric configurations used an increased temperature gradient of $a / L_{T s} = 3.0$. The effect of this was measured by performing a mirror symmetric simulation at both $a / L_{T s} = 2.3$ and $a / L_{T s} = 3.0$. This change was found to modify the ratio of the momentum flux to the heat flux by less than $5\%$. We are interested in this ratio because we expect it to be roughly proportional to the level of intrinsic rotation \cite{BallShafranovShift2016}.

\begin{figure}
 \centering
 \includegraphics[width=0.8\textwidth]{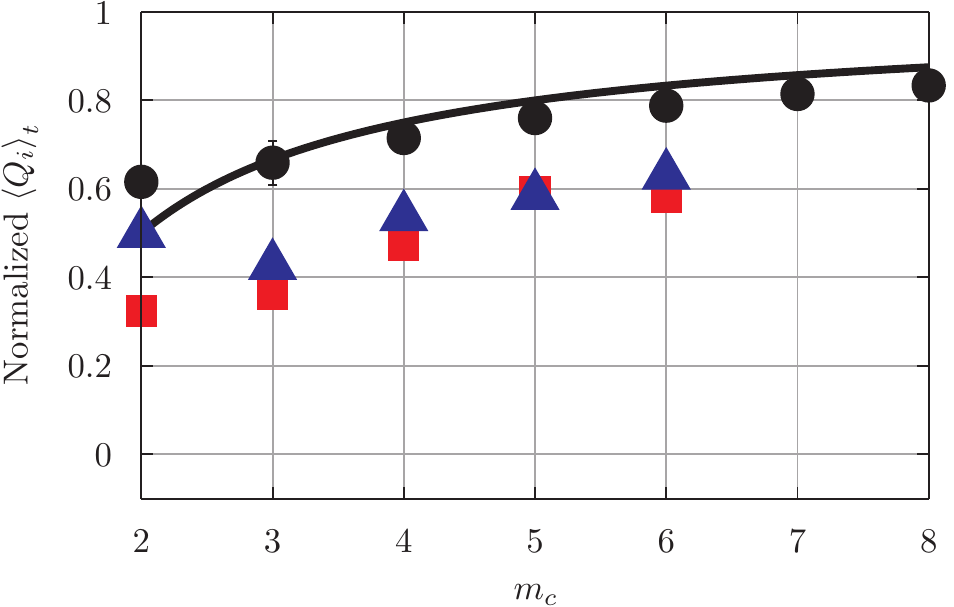}
 \caption{The time-averaged ion energy flux from mirror symmetric flux surfaces (black, circles), non-mirror symmetric flux surfaces with up-down symmetric envelopes (red, squares), and non-mirror symmetric flux surfaces with up-down asymmetric envelopes (blue, triangles) all normalized to the value from a circular flux surface. Also shown is the $m_{c}^{-1}$ scaling (black, solid) expected for all three scans and a single set of error bars representative of the error in all data points.}
 \label{fig:energyFlux}
\end{figure}

Figure \ref{fig:energyFlux} shows the time-averaged radial flux of energy carried by the ions $\left\langle Q_{i} \right\rangle_{t}$, calculated by GS2 for the three scans. Like the momentum flux, we expect the energy flux to approach the value of a circular flux surface in the $m_{c} \gg 1$ limit. However, unlike the momentum flux, the energy flux does not cancel on circular flux surfaces, so this value is not zero. Furthermore, we expect all three scan to tend to the circular value as $m_{c}^{-1}$. This is because there are $\Order{m_{c}^{-1}}$ terms in the geometric coefficients that do not disappear after averaging over $z$ (e.g. the first term in equation \refEq{eq:nonMirrorAlphaDriftReworked}). These do not contribute to the momentum transport because they are up-down symmetric, but they still modify the energy flux. These theoretical expectations are all consistent with the numerical data in figure \ref{fig:energyFlux}.

\begin{figure}
 \centering
 \includegraphics[width=0.8\textwidth]{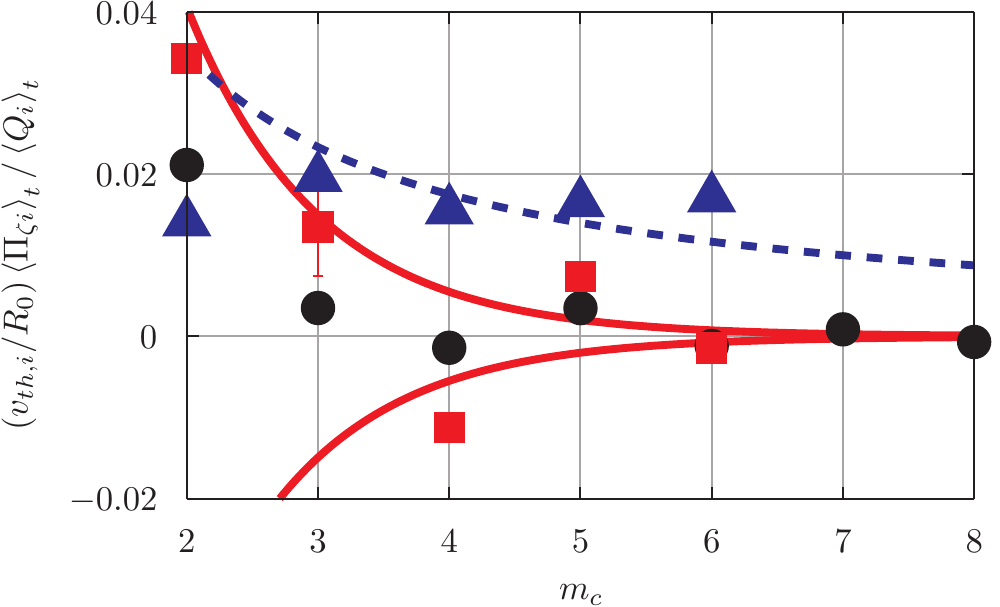}
 \caption{The time-averaged ion momentum flux from mirror symmetric flux surfaces (black, circles), non-mirror symmetric flux surfaces with up-down symmetric envelopes (red, squares), and non-mirror symmetric flux surfaces with up-down asymmetric envelopes (blue, triangles). Also shown is the $m_{c}^{-1}$ scaling expected for the scan with up-down asymmetric envelopes (blue, dotted), an $\Exp{-m_{c}}$ scaling that may be followed by the other two scans (red, solid), and a single set of error bars representative of the error in all data points. Here $v_{th, i}$ is the ion thermal velocity.}
 \label{fig:momHeatFluxRatio}
\end{figure}

Figure \ref{fig:momHeatFluxRatio} shows the time-averaged radial flux of ion toroidal angular momentum $\left\langle \Pi_{\zeta i} \right\rangle_{t}$. We see behavior that is consistent with the analytic scalings from section \ref{sec:analytics}. We expect both the mirror symmetric flux surfaces and non-mirror symmetric flux surfaces with up-down symmetric envelopes to have momentum transport that decreases exponentially. Conversely, we expect the non-mirror symmetric flux surfaces with up-down asymmetric envelopes to have momentum transport that only decreases polynomially. Accordingly, we see that the momentum flux in the scan with the asymmetric envelope decreases much more slowly than the other two scans and has significantly higher values at large $m_{c}$.

\section{Conclusions}
\label{sec:conclusions}

In this work, we used analytic and numerical results to conclude that high-order flux surface shaping generally drives momentum transport that is polynomially small in the mode number of the high-order shaping. However, when the high-order shaping is mirror symmetric and does not create an up-down asymmetric slowly-varying envelope, the momentum transport must be exponentially small. This theoretical conclusion indicates two concrete, experimentally-relevant points. First, using low-order shaping effects to break the up-down symmetry of flux surfaces will lead to faster intrinsic rotation. This is good as low-order shaping is easier to create with external coils and better penetrates from the plasma edge into the core \cite{BallMastersThesis2013, RodriguesMHDupDownAsym2014, BizarroUpDownAsymGradShafEq2014, BallShapingPenetration2015}. Second, this work suggests that it is important to explore (both computationally and experimentally) non-mirror symmetric shapes as well as shapes with up-down asymmetric envelopes. These unusual shapes have the potential to significantly increase the level of intrinsic rotation generated by up-down asymmetry.

Flux surfaces that are non-mirror symmetric and have an up-down asymmetric envelope can be created in current experiments. However, doing so may require a reduction in the plasma volume and/or the plasma current (among other considerations). This is because the shape of the vacuum vessel is typically designed to closely fit a conventional up-down symmetric plasma shape. Therefore, by simple geometry, changing the shape without touching the vacuum vessel necessitates a reduction in volume. Additionally, the locations and current limits of the shaping coils were chosen with conventional flux surface shapes in mind. Creating a significantly different shape may involve violating a current limit in one or more of the shaping coils. This can be resolved by uniformly reducing all of the currents that contribute to the poloidal field, which includes the plasma current.

Hence, in current experiments making significant changes to the plasma shape likely involves a reduction in the plasma volume/current. The amount depends on the specifics of the machine as well as the desired shape. Therefore, for a given experiment we can use free-boundary MHD equilibrium analysis together with gyrokinetic turbulence simulations to identify specific shapes that require little reduction in the plasma volume/current, yet still drive a lot of rotation. Lastly, we note that this reduction in plasma parameters is a consequence of creating a plasma shape that current experiments were not optimized to create. A device that was designed to create a specific plasma shape that is non-mirror symmetric and has an up-down asymmetric envelope would not encounter these problems. Thus, we can use current experiments to investigate the performance of these unusual shapes and, if one of them proves exceptional, we know it can be implemented more effectively in a dedicated experiment.

\ack

This work was funded in part by the RCUK Energy Programme (grant number EP/I501045). Computing time for this work was provided by the Helios supercomputer at IFERC-CSC under the projects SPIN, TRIN, MULTEIM, and GKMSC. The authors also acknowledge the use of ARCHER through the Plasma HEC Consortium EPSRC grant number EP/L000237/1 under the projects e281-gs2 and e281-rotation.

\appendix

\section{Analytic calculation of the envelope}
\label{app:envCalc}

The slowly-varying envelope $r_{0, \text{env}} \left( \theta \right)$ arises from the family of flux surface shapes generated by changing $z_{t}$ in $r_{0} \left( \theta, z + z_{t} \right)$. The envelope is specified by equation \refEq{eq:env}, where the key function $z_{\text{env}} \left( \theta \right)$ is calculated by enforcing equation \refEq{eq:envParameter}. In this appendix we will consider a flux surface shaped by two Fourier modes $m$ and $n$ that beat against one another to generate an envelope. We will choose the characteristic mode number to be $m_{c} = m$ and assume that $n$ is sufficiently close that both modes map to the same value of $l = 1$ (according to equation \refEq{eq:lDef}). With this geometry specification, we will first calculate $z_{\text{env}} \left( \theta \right)$, then find an analytic expression for the envelope, and finally derive the mode tilt angles for which the envelope is up-down symmetric.

We start with the general flux surface specification, given by equation \refEq{eq:fluxSurfaceSpecScaleSep}, and include only two shaping modes. Substituting this specification shows that
\begin{align}
   r_{0} \left( \theta, z_{\text{env}} \left( \theta \right) \right) = r_{\psi 0} \bigg[ 1 &- \frac{\Delta_{m} - 1}{\Delta_{m} + 1} \Cos{z_{\text{env}} \left( \theta \right) + m \theta_{t m}} \label{eq:twoModeGeoSpec} \\
   &- \frac{\Delta_{n} - 1}{\Delta_{n} + 1} \Cos{z_{\text{env}} \left( \theta \right) + \left( n - m \right) \theta + n \theta_{t n}} \bigg] , \nonumber
\end{align}
where we still must determine the unknown function $z_{\text{env}} \left( \theta \right)$. To calculate it we follow equation \refEq{eq:envParameter} and take the derivative of \refEq{eq:twoModeGeoSpec} to find
\begin{align}
   \frac{\Delta_{m} - 1}{\Delta_{m} + 1} &\Sin{z_{\text{env}} \left( \theta \right) + m \theta_{t m}} + \frac{\Delta_{n} - 1}{\Delta_{n} + 1} \nonumber \\
   \times \big[ &\Sin{z_{\text{env}} \left( \theta \right) + m \theta_{t m}} \Cos{\left( n - m \right) \theta + n \theta_{t n} - m \theta_{t m}} \label{eq:simplifiedZtCond} \\
   + &\Cos{z_{\text{env}} \left( \theta \right) + m \theta_{t m}} \Sin{\left( n - m \right) \theta + n \theta_{t n} - m \theta_{t m}} \big] = 0 . \nonumber
\end{align}
By dividing by $\Cos{z_{\text{env}} \left( \theta \right) + m \theta_{t m}}$, we see that this equation is solved by
\begin{align}
   z_{\text{env}} \left( \theta \right) = K_{2} \pi - m \theta_{t m} - \ArcTan{\frac{\frac{\Delta_{n} - 1}{\Delta_{n} + 1} \Sin{\left( n - m \right) \theta + n \theta_{t n} - m \theta_{t m}}}{{\frac{\Delta_{m} - 1}{\Delta_{m} + 1} + \frac{\Delta_{n} - 1}{\Delta_{n} + 1} \Cos{\left( n - m \right) \theta + n \theta_{t n} - m \theta_{t m}}}}} , \label{eq:ztSol}
\end{align}
where $K_{2}$ is a free integer that selects the inner or outer envelope. Combining equations \refEq{eq:env}, \refEq{eq:twoModeGeoSpec}, and \refEq{eq:ztSol} and simplifying gives
\begin{align}
   r_{0, \text{env}} &\left( \theta \right) = r_{\psi 0} \Bigg( 1 - \left( -1 \right)^{K_{2}} \label{eq:envSol} \\
   \times& \sqrt{\left( \frac{\Delta_{m} - 1}{\Delta_{m} + 1} \right)^{2} + \left( \frac{\Delta_{n} - 1}{\Delta_{n} + 1} \right)^{2} + 2 \frac{\Delta_{m} - 1}{\Delta_{m} + 1} \frac{\Delta_{n} - 1}{\Delta_{n} + 1} \Cos{\left( n - m \right) \theta + n \theta_{t n} - m \theta_{t m}}} \Bigg) . \nonumber
\end{align}
This is the general solution for the envelope formed by the beating of two modes. We know that the envelope will be up-down symmetric if
\begin{align}
   r_{0, \text{env}} \left( \theta \right) = r_{0, \text{env}} \left( - \theta \right) ,
\end{align}
as long as $\theta$ is defined such that $\theta = 0$ is on the midplane (as is the case in this work). For this relationship to be satisfied given equation \refEq{eq:envSol}, we require that
\begin{align}
   n \theta_{t n} - m \theta_{t m} = K_{1} \pi ,
\end{align}
where $K_{1}$ is an integer. This condition on the mode tilt angles, which is equivalent to equation \refEq{eq:twoModeEnvelopeCond}, is satisfied if and only the envelope of the flux surface is up-down symmetric.

\section{Geometric coefficients appearing in gyrokinetics}
\label{app:nonMirrorGeoCoeffs}

The $O \left( 1 \right)$ geometric coefficients to lowest order in aspect ratio are simply those of a circular tokamak and are given by
\begin{align}
  \left( \hat{b} \cdot \Nabla \theta \right)_{0} &= \frac{1}{r_{\psi 0} R_{0} B_{0}} \frac{d \psi}{d r_{\psi}} \label{eq:gradparO0} \\
  \left( \vec{v}_{d s} \cdot \Nabla \psi \right)_{0} &= - \frac{1}{R_{0} \Omega_{s}} \frac{d \psi}{d r_{\psi}} \Sin{\theta} \label{eq:psiDriftO0} \\
  \left( \vec{v}_{d s} \cdot \Nabla \alpha \right)_{0} &= \frac{B_{0}}{R_{0} \Omega_{s}} \left( \frac{d \psi}{d r_{\psi}} \right)^{-1} \left( \Cos{\theta} + \hat{s}' \theta \Sin{\theta} \right) \label{eq:alphaDriftO0} \\
  \left| \Nabla \psi \right|^{2}_{0} &= \left( \frac{d \psi}{d r_{\psi}} \right)^{2} \\
  \left( \Nabla \psi \cdot \Nabla \alpha \right)_{0} &= - B_{0} \hat{s}' \theta \\
  \left| \Nabla \alpha \right|^{2}_{0} &= B_{0}^{2} \left( \frac{d \psi}{d r_{\psi}} \right)^{-2} \left( 1 + \hat{s}'^{2} \theta^{2} \right) \label{eq:gradAlphaSqO0} \\
  \left( J_{0} \left( k_{\perp} \rho_{s} \right) \right)_{0} &= J_{0} \left( k_{\perp 0} \rho_{s} \right) , \label{eq:FLRO0}
\end{align}
where
\begin{align}
  k_{\perp 0} \rho_{s} &\equiv \rho_{s} \sqrt{k_{\psi}^{2} \left| \Nabla \psi \right|^{2}_{0} + 2 k_{\psi} k_{\alpha} \left( \Nabla \psi \cdot \Nabla \alpha \right)_{0} + k_{\alpha}^{2} \left| \Nabla \alpha \right|^{2}_{0}} . \label{eq:FLR0def}
\end{align}
Here
\begin{align}
  \hat{s}' &\equiv 2 + r_{\psi 0} \left( \frac{d \psi}{d r_{\psi}} \right)^{-1} \left( \mu_{0} R_{0}^{2} \frac{d p}{d \psi} + R_{0} B_{0} \frac{d I}{d \psi} \right) , \label{eq:shiftedShatDef}
\end{align}
$J_{n} \left( \ldots \right)$ is the $n$th order Bessel function of the first kind, $\rho_{s}$ is the gyroradius, $k_{\psi}$ is the radial wavenumber of the turbulence, $k_{\alpha}$ is the poloidal wavenumber of the turbulence, $\mu_{0}$ is the vacuum permeability, and $p$ is the plasma pressure. Note that all of the coefficients are independent of the short spatial scale coordinate, $z$.

To $O \left( m_{c}^{-1} \right)$ the geometric coefficients to lowest order in aspect ratio are
\begin{align}
  \left( \hat{b} \cdot \Nabla \theta \right)_{1} &= \frac{1}{2 R_{0} B_{0}} \frac{d \psi}{d r_{\psi}} \left( \frac{d \Delta_{m}}{d r_{\psi}} \Cos{z_{m s}} + \frac{d \Delta_{n}}{d r_{\psi}} \Cos{z_{n s}} \right) \label{eq:gradparO1}
\end{align}
\begin{align}
  \left( \vec{v}_{d s} \cdot \Nabla \psi \right)_{1} &= \frac{1}{2 R_{0} \Omega_{s}} \frac{d \psi}{d r_{\psi}} \Bigg[ \Cos{\theta} \Big( m \left( \Delta_{m} - 1 \right) \Sin{z_{m s}} + n \left( \Delta_{n} - 1 \right) \Sin{z_{n s}} \Big) \nonumber \\
  &- r_{\psi 0} \Sin{\theta} \left( \frac{d \Delta_{m}}{d r_{\psi}} \Cos{z_{m s}} + \frac{d \Delta_{n}}{d r_{\psi}} \Cos{z_{n s}} \right) \Bigg] \label{eq:psiDriftO1}
\end{align}
\begin{align}
  \left( \vec{v}_{d s} \cdot \Nabla \alpha \right)_{1} &= \frac{B_{0}}{2 R_{0} \Omega_{s}} \left( \frac{d \psi}{d r_{\psi}} \right)^{-1} \nonumber \\
  \times& \Bigg[ r_{\psi 0} \left( m^{2} \left( \Delta_{m} - 1 \right) \frac{d \Delta_{m}}{d r_{\psi}} + n^{2} \left( \Delta_{n} - 1 \right) \frac{d \Delta_{n}}{d r_{\psi}} \right) \theta \Sin{\theta} \nonumber \\
  -& \Big( \Sin{\theta} + \hat{s}' \theta \Cos{\theta} \Big) \Big( m \left( \Delta_{m} - 1 \right) \Sin{z_{m s}} + n \left( \Delta_{n} - 1 \right) \Sin{z_{n s}} \Big) \nonumber \\
  -& r_{\psi 0} \Big( \Cos{\theta} - \hat{s}' \theta \Sin{\theta} \Big) \left( \frac{d \Delta_{m}}{d r_{\psi}} \Cos{z_{m s}} + \frac{d \Delta_{n}}{d r_{\psi}} \Cos{z_{n s}} \right) \label{eq:alphaDriftO1} \\
  +& \frac{r_{\psi 0}}{\left( n - m \right)} \left( m^{2} \left( \Delta_{m} - 1 \right) \frac{d \Delta_{n}}{d r_{\psi}} + n^{2} \left( \Delta_{n} - 1 \right) \frac{d \Delta_{m}}{d r_{\psi}} \right) \Sin{\theta} \nonumber \\
  \times& \Big( \Sin{\left( n - m \right) \theta} \Cos{m \theta_{t m} - n \theta_{t n}} \nonumber \\
  &- \Cos{\left( n - m \right) \theta} \Sin{m \theta_{t m} - n \theta_{t n}} \Big) \Bigg] \nonumber
\end{align}
\begin{align}
  \left| \Nabla \psi \right|^{2}_{1} &= r_{\psi 0} \left( \frac{d \psi}{d r_{\psi}} \right)^{2} \left( \frac{d \Delta_{m}}{d r_{\psi}} \Cos{z_{m s}} + \frac{d \Delta_{n}}{d r_{\psi}} \Cos{z_{n s}} \right) \label{eq:gradPsiSqO1}
\end{align}
\begin{align}
  \left( \Nabla \psi \cdot \Nabla \alpha \right)_{1} &= - B_{0} \Bigg[ \frac{r_{\psi 0}}{2} \left( m^{2} \left( \Delta_{m} - 1 \right) \frac{d \Delta_{m}}{d r_{\psi}} + n^{2} \left( \Delta_{n} - 1 \right) \frac{d \Delta_{n}}{d r_{\psi}} \right) \theta \nonumber \\
  -& \Big( m \left( \Delta_{m} - 1 \right) \Sin{z_{m s}} + n \left( \Delta_{n} - 1 \right) \Sin{z_{n s}} \Big) \nonumber \\
  +& r_{\psi 0} \hat{s}' \theta \left( \frac{d \Delta_{m}}{d r_{\psi}} \Cos{z_{m s}} + \frac{d \Delta_{n}}{d r_{\psi}} \Cos{z_{n s}} \right) \label{eq:gradPsiDotGradAlphaO1} \\
  +& \frac{r_{\psi 0}}{2 \left( n - m \right)} \left( m^{2} \left( \Delta_{m} - 1 \right) \frac{d \Delta_{n}}{d r_{\psi}} + n^{2} \left( \Delta_{n} - 1 \right) \frac{d \Delta_{m}}{d r_{\psi}} \right) \nonumber \\
  \times& \Big( \Sin{\left( n - m \right) \theta} \Cos{m \theta_{t m} - n \theta_{t n}} \nonumber \\
  &- \Cos{\left( n - m \right) \theta} \Sin{m \theta_{t m} - n \theta_{t n}} \Big) \Bigg] \nonumber
\end{align}
\begin{align}
  \left| \Nabla \alpha \right|^{2}_{1} &= B_{0}^{2} \left( \frac{d \psi}{d r_{\psi}} \right)^{-2} \Bigg[ r_{\psi 0} \left( m^{2} \left( \Delta_{m} - 1 \right) \frac{d \Delta_{m}}{d r_{\psi}} + n^{2} \left( \Delta_{n} - 1 \right) \frac{d \Delta_{n}}{d r_{\psi}} \right) \hat{s}' \theta^{2} \nonumber \\
  -& 2 \hat{s}' \theta \Big( m \left( \Delta_{m} - 1 \right) \Sin{z_{m s}} + n \left( \Delta_{n} - 1 \right) \Sin{z_{n s}} \Big) \nonumber \\
  -& r_{\psi 0} \left( 1 - \hat{s}'^{2} \theta^{2} \right) \left( \frac{d \Delta_{m}}{d r_{\psi}} \Cos{z_{m s}} + \frac{d \Delta_{n}}{d r_{\psi}} \Cos{z_{n s}} \right) \label{eq:gradAlphaSqO1} \\
  +& \frac{r_{\psi 0}}{n - m} \left( m^{2} \left( \Delta_{m} - 1 \right) \frac{d \Delta_{n}}{d r_{\psi}} + n^{2} \left( \Delta_{n} - 1 \right) \frac{d \Delta_{m}}{d r_{\psi}} \right) \hat{s}' \theta \nonumber \\
  \times& \Big( \Sin{\left( n - m \right) \theta} \Cos{m \theta_{t m} - n \theta_{t n}} \nonumber \\
  &- \Cos{\left( n - m \right) \theta} \Sin{m \theta_{t m} - n \theta_{t n}} \Big) \Bigg] \nonumber
\end{align}
\begin{align}
  \left( J_{0} \left( k_{\perp} \rho_{s} \right) \right)_{1} &= - k_{\perp 1} \rho_{s} J_{1} \left( k_{\perp 0} \rho_{s} \right) , \label{eq:FLRO1avg}
\end{align}
where
\begin{align}
  k_{\perp 1} \rho_{s} &\equiv \frac{k_{\perp 0} \rho_{s}}{2} \frac{k_{\psi}^{2} \left| \Nabla \psi \right|_{1}^{2} + 2 k_{\psi} k_{\alpha} \left( \Nabla \psi \cdot \Nabla \alpha \right)_{1} + k_{\alpha}^{2} \left| \Nabla \alpha \right|_{1}^{2}}{k_{\psi}^{2} \left| \Nabla \psi \right|_{0}^{2} + 2 k_{\psi} k_{\alpha} \left( \Nabla \psi \cdot \Nabla \alpha \right)_{0} + k_{\alpha}^{2} \left| \Nabla \alpha \right|_{0}^{2}} . \label{eq:FLRO1def}
\end{align}
From the last terms in each of \refEq{eq:alphaDriftO1}, \refEq{eq:gradPsiDotGradAlphaO1}, and \refEq{eq:gradAlphaSqO1} we see that (even after averaging over $z$) $\left( \vec{v}_{d s} \cdot \Nabla \alpha \right)_{1}$, $\left( \Nabla \psi \cdot \Nabla \alpha \right)_{1}$, and $\left| \Nabla \alpha \right|^{2}_{1}$ are all up-down asymmetric.

\section*{References}
\bibliographystyle{unsrt}
\bibliography{references.bib}

\end{document}